\documentclass{emulateapj}

\usepackage{epsf}
\usepackage{graphics}
\usepackage{color}

\newcommand{\be}{\begin{equation}}
\newcommand{\ee}{\end{equation}}

\newcommand{\kms}{km s$^{-1}$}

\newcommand{\ax}{$\alpha_{\rm X}$}
\newcommand{\axs}{$\alpha_{\rm X, soft}$}
\newcommand{\axh}{$\alpha_{\rm X, hard}$}
\newcommand{\aox}{$\alpha_{\rm ox}$}

\newcommand{\rb}[1]{\raisebox{1.5ex}[-1.5ex]{#1}}
\newcommand{\msun}{$M_{\odot}$}

\newcommand{\plm}{$\pm$}
\newcommand{\nh}{$N_{\rm H}$}
\newcommand{\swift}{{\it Swift}}
\newcommand{\rxj}{{\rm RX\,J0148.3--2758}}

\shorttitle{Swift observation of RX J0148.3--2758}
\shortauthors{Grupe et al.}

\begin{document}

%\input DGrupe_clipfig.tex
%\useunitmm

\def\etal{{\it et\thinspace al.}\ }
\def\alp{{$\alpha$}\ }
\def\al2{{$\alpha^2$}\ }

%\def \charthoffset {\hspace{0.2cm}} \def \charthsep {\hspace{0.3cm}}
%\def \chartvsepcap {\vspace{0.3cm}}
%\def \chartvsep {\vspace{0.1cm}}
%\newcommand{\putchartb}[1]{\clipfig{#1}{75}{20}{7}{275}{192}}
%\newcommand{\putchartc}[1]{\clipfig{#1}{55}{33}{19}{275}{195}}
%
%
%
%\newcommand{\chartlineb}[2]{\parbox[t]{18cm}{\noindent\charthoffset\putchartb{#1}\charthsep\putchartb{#2}\chartvsep}}
%\newcommand{\chartlinec}[3]{\parbox[t]{18cm}{\noindent\charthoffset\putchartc{#1}\charthsep\putchartc{#2}\chartvsep\putchartc{#3}\chartvsep}}

%% LaTeX will automatically break titles if they run longer than
%% one line. However, you may use \\ to force a line break if
%% you desire.

\title{Swift Observations of the highly X-ray variable Narrow Line Seyfert 1 galaxy
\rxj
}

%% Use \author, \affil, and the \and command to format
%% author and affiliation information.
%% Note that \email has replaced the old \authoremail command
%% from AASTeX v4.0. You can use \email to mark an email address
%% anywhere in the paper, not just in the front matter.
%% As in the title, you can use \\ to force line breaks.

\author{Dirk Grupe\altaffilmark{1}
\email{grupe@astro.psu.edu},
Karen M. Leighly\altaffilmark{2},
Stefanie Komossa\altaffilmark{3},
Patricia Schady\altaffilmark{1,4},
Paul T. O'Brien\altaffilmark{5},
David N. Burrows\altaffilmark{1}, John A. Nousek\altaffilmark{1}
\email{dxb15@psu.edu, nousek@astro.psu.edu} 
}

\altaffiltext{1}{Department of Astronomy and Astrophysics, Pennsylvania State
University, 525 Davey Lab, University Park, PA 16802} 

\altaffiltext{2}{Homer L. Dodge Department of Physics and Astronomy, 
University of Oklahoma, 
440 West Brooks Street, Norman, OK 73019; email: leighly@nhn.ou.edu}

\altaffiltext{3}{MPI f\"ur extraterrestrische Physik, Giessenbachstr., D-85748 Garching,
Germany; email: skomossa@mpe.mpg.de}

\altaffiltext{4}{Mullard Space Science Laboratory, Holmbury St. Mary, Dorking,
Surrey RH5 6NT, U.K.; 
email: ps@mssl.ucl.ac.uk}

\altaffiltext{5}{Department of Physics \& Astronomy, University of Leicester,
Leicester, LE1 7R, UK, email: pto@star.le.ac.uk}

%% Notice that each of these authors has alternate affiliations, which
%% are identified by the \altaffilmark after each name.  Specify alternate
%% affiliation information with \altaffiltext, with one command per each
%% affiliation.

%\altaffiltext{1}{Visiting Astronomer, Cerro Tololo Inter-American Observat}

%% Mark off your abstract in the ``abstract'' environment. In the manuscript
%% style, abstract will output a Received/Accepted line after the
%% title and affiliation information. No date will appear since the author
%% does not have this information. The dates will be filled in by the
%% editorial office after submission.

\begin{abstract}
We report on  \swift\ observations of the  Narrow-Line Seyfert 1 galaxy 
(NLS1) RX J0148.3--2758. It was observed for 41.6 ks in 2005 May and for 15.8 ks in 2005 December.
On short as well as
on long timescales RX J0148.3--2758 is a highly variable source. It doubles 
its X-ray flux within 18-25 ks.
The observation of 2005 December 09, which had a flux 4 times lower than during the 2005 May
observations,
shows a significant hardening of the X-ray hardness ratio compared with the 2005-May
and 2005-December 20/21 observations. A detailed analysis of the X-ray spectra shows that
 we actually observe two spectral changes in \rxj: first, a decrease of the soft X-ray component
 between 2005 May and December 09, which is most likely due to an increase of the intrinsic
 absorber column, and second, a decrease of the hard X-ray flux in the December 20/21
 observations.
 The soft X-ray spectral slope 
$\alpha_{\rm X, soft}$=2.58$^{+0.15}_{-0.12}$ during the high state in 2005 May 
agrees well with that measured by {\it ROSAT} (\axs=2.54\plm0.82). This soft X-ray
spectrum is superimposed on a hard
X-ray component with $\alpha_{\rm X, hard}$=0.96$^{+0.15}_{-0.12}$ which in
consistent with the hard X-ray spectral slope \axh=1.11$^{+0.16}_{-0.19}$ found by {\it ASCA}.
The soft X-ray slope \axs=1.93$^{+0.58}_{-0.42}$ measured during the December 09
observation, agrees well with \axs=2.03$^{+0.23}_{-0.20}$ measured from the {\it ASCA}
observation when \rxj~ was also in a low state.
In contrast to the strong X-ray variability, 
the analysis of the \swift~UVOT photometry from December 2005 of RX J0148.3--2758 shows 
no significant  variability in any of the 6 UVOT filters. 
From the simultaneous X-ray and UV observations in 2005 December we measured the X-ray
loudness \aox~ and found it to vary between \aox=1.5 and 1.8.
Our \swift\ observations of \rxj~
demonstrate the great potential that the multi-wavelength observatory
\swift\ has for AGN science. 
\end{abstract}


\section{Introduction}

Most of the power in the spectral energy distribution (SED) of an AGN
is contained in the Big Blue Bump \citep[BBB, ][]{shields78}. As
suggested by \citet{walter93}, the BBB may stretch from the UV into
the soft X-ray regime. The soft X-ray part of the BBB may be UV
photons from the accretion disk which are shifted into the soft X-ray
band by Comptonization in the accretion disk corona
\citep[e.g. ][]{pounds95}. Based on their sample of soft X-ray
selected {\it ROSAT} AGN, \citet{gru98a} showed that the BBB extends as far
as the optical band and that sources with steeper X-ray spectra tend to
have bluer optical spectra, suggesting that Narrow Line Seyfert 1
galaxies are the AGN with the strongest BBB component. However, from a
study of the IUE spectra of NLS1s, \citet{rodriguez97} came to the
conclusion that NLS1s have weaker UV emission than Broad Line Seyfert
1s.  All these studies, however, were hampered by the lack of
simultaneous observations in the optical/UV and X-ray bands; the
observations available frequently had been 
performed years apart. This situation has changed now with the
availability of the multi-wavelengths observatories XMM-Newton and
\swift.

\swift\ \citep{gehrels04} is a multi-wavelength mission equipped with
three telescopes that together cover the electromagnetic spectrum
between 6000\AA~to 150 keV: the Burst Alert Telescope \citep[BAT,
][]{barthelmy04}, the X-Ray Telescope \citep[XRT, ][]{burrows04}, and
the UV-Optical Telescope \citep[UVOT, ][]{roming04}. \swift\ was
designed to chase Gamma-Ray Bursts (GRBs), and was launched on
20-November-2004.  At the low-energy side of \swift's observing
window, the UVOT covers the wavelengths range between 1700-6000\AA.
The UVOT is a sister instrument of XMM's Optical Monitor \citep[OM,
][]{mason01}, equipped with a similar set of filters \citep{mason01,
roming04}. The XRT covers the 0.3-10.0 keV range, and  uses a CCD
detector identical to the EPIC MOS on-board XMM 
\citep{tur01}. As described by \citet{hill04} the XRT operates in
three observing modes: the Photon Counting (PC) which is equivalent to
the full-frame mode on XMM, Window Timing (WT), and Low-Rate
Photo-diode mode (LrPD). Due to the nature of the \swift~mission, the
XRT switches automatically between the observing modes according to
the brightness of a source. Only for specific purposes, e.g. in case
of calibration observations, are the modes set manually in the
observing schedule. The BAT is a coded-mask experiment that operates
in the 15-150 keV energy range, at the high-energy part of \swift's
observing window. Although the main purpose of the {\it Swift} mission
is to detect and observe GRBs, fill-in targets are used in the
observing schedule to optimize the scientific return of the mission
when GRBs are not observable.  \swift\'s UV and X-ray capabilities,
and rapid and flexible scheduling make it an ideal observatory to
study AGN.

With the launch of the X-ray satellite {\it ROSAT} \citep{tru83} the X-ray
energy range down to 0.1 keV became accessible for the first
time. During the half-year {\it ROSAT} All-Sky Survey (RASS, \citep[RASS,
][]{vog99} a large number of sources with steep X-ray spectra were
detected \citep{tho98, beu99, schwo00}.  About one third to one half
of these sources are AGN. \citet{gru96} and \citet{gru98a, gru04a}
found that about 50\% of bright soft X-ray selected AGN are
Narrow-Line Seyfert 1 galaxies \citep[NLS1s,][]{oster85, good89}. They
turned out to be the class of AGN with the steepest X-ray spectra
\citep[e.g.,][]{puch92, bol96, gru96, gru98a, gru01, vau01, gru04a,
williams02} and often show very strong X-ray variability
\citep[e.g.,][]{bol96, nandra97, lei99, turner99, gru01}.  NLS1s are
AGN with extreme properties which seem to be linked to one another: a
steeper X-ray spectral index \ax~correlates with the
strength of the optical FeII emission and anti-correlates with the
widths of the Broad Line Region (BLR) Balmer lines and the strength of
the Narrow-Line Region (NLR) forbidden lines \citep[e.g.,][]{gru96,
gru99, gru03b, laor94, laor97, sul00}.  All these relationships are
governed by a set of fundamental underlying parameters, usually called
the \citet{bor92} 'Eigenvector-1' relation in AGN. The most accepted
explanation for these Eigenvectors is the Eddington ratio $L/L_{Edd}$
or the mass of the central black hole $M_{\rm BH}$ \citep{bor02,
sul00, gru03b, yuan03} in which NLS1s are AGN with the highest
Eddington ratios and smallest black hole masses for a given
luminosity.  The Eddington ratio has also been found to be correlated
with the X-ray spectral slope \ax~\citep{gru03b}.  Alternatively, this
can also be interpreted as the age of an AGN in which NLS1s are AGN in
an early stage of their development \citep{gru96, gru03b, mat00}.

RX J0148.3--2758 ($\alpha_{2000}$=01 48 22.3, $\delta_{2000}$=--27 58
26, z=0.121) was discovered during the RASS as a bright and variable
X-ray source \citep{gru98a,tho98,schwo00}. It was identified as a NLS1
by \citet{gru96} and \citet{gru99}.  Besides a later 6.7 ks {\it ROSAT} PSPC
observation \citep{gru01}, RX J0148.3--2758 was also observed for 34
ks by {\it ASCA} \citep{turner99, vau99}.  The 2-10 keV {\it ASCA} light curve
shows that the source is highly variable \citep{turner99}. Its 2-10
keV spectral slope $\alpha_{\rm 2-10~keV}$=0.99\plm0.17 is typical for
a Seyfert 1 galaxy \citep{vau99}. In this paper we present our
observations of RX J0148.3--2758 with \swift\, and we compare those
with the data previously taken by {\it ROSAT} and {\it ASCA}. \rxj\ was one of the
most X-ray variable AGN in the soft X-ray selected AGN sample of
\citet{gru01}. 

The outline of this paper is as follows: in \S\,1
%\ref{observe} 
we
describe the \swift, {\it ROSAT} and {\it ASCA} observations and the data
reduction, in \S\,3
%\ref{results} 
we present the results of the \swift\
data analysis, and in \S\,4
%\ref{discuss} 
we discuss the results.
Throughout the paper, spectral indexes are quoted as energy spectral
indexes with $F_{\nu} \propto \nu^{-\alpha}$. Luminosities are
calculated assuming a $\Lambda$CDM cosmology with $\Omega_{\rm
M}$=0.27, $\Omega_{\Lambda}$=0.73 and a Hubble constant of $H_0$=75 km
s$^{-1}$ Mpc$^{-1}$ using the luminosity distances given by
\citet{hogg99}. All errors are 1$\sigma$ unless stated otherwise.

\section{\label{observe} Observations and data reduction}

RX J0148.3--2758 was observed by \swift\ between 2005-May-05 and
2005-May-13 (segments 002-006) for a total of 41.6 ks and between
2005-December-07 and 2005-December-21 for 15.8 ks (segments 008-011).
Table \ref{obs_log} lists the segment numbers of the \swift\
observations, the start and end times, the total observing times and
the 0.2-2.0 keV rest-frame luminosities.  All XRT observations were
performed in PC mode. The event files were created with the standard
Swift XRT analysis task {\it xrtpipeline} version 0.9.9.  For both
spectral and temporal analysis, source counts  between
0.3-10 keV were extracted from a circle with a radius of 50$^{''}$,
and background photons were extracted from a 100$^{''}$ circle in a
source-free region near the source.  We created source and background
spectra and 
event files using {\it XSELECT} version 2.3.  Background-subtracted
light curves were created by using ESO's Munich Image Data Analysis
System MIDAS version 04Sep as described in \citet{nousek06}.  The data
were binned to have 250 source + background photons per bin except for
the 2005 December 07 observation where we used a binning of 150
photons per bin. Note that on 2005 May 27 the \swift\ XRT detector was
hit by a micro-meteorite that caused some damages. In particular the
CCD columns DETX=294 and 320 had to be turned off afterwards. While
our 2005 May and the 2005 December 20/21 observations are not affected
by those dead columns, the 2005 December 07 and 09 observations were
in part.  For the latter data sets, a correction was applied if the
source felt on one of the dead columns to account for the loss of
photons this caused.  The spectra were rebinned using {\it grppha}
version 3.0.0 to have at least 20 photons per bin and analyzed using
{\it XSPEC} 12.2.1 \citep{arnaud96}. 
The Auxiliary Response files were created using
the \swift\ analysis task {\it xrtmkarf}. We used the standard
response matrix version 007 with grade selection 0 to 12.  Due to the
low count rate, the data were not affected by pileup.
 
\swift\ UVOT data were obtained during 2005 May 11 and 13 (segments
004 and 006), and 2005 December.  The UVOT was blocked during 2005 May
05 and 07 (segments 002 and 004) observations.  The UV grism was used for the
observations of 2005 May 11 and 13, and during 2005 December UVOT
photometry was performed.  Due to the on-going calibration of the UV
grisms and the requirement for well-calibrated UV grism data in our
analysis, we do not present these data at this point, and discuss only
the UVOT photometry results of the 2005 December observations.

Observations were taken in the three optical and three UV filters
 available on the UVOT \citep{roming04} with the exception of 2005
 December 07 (segment 008) observation, in which no B band
 observations were  made. This covers the wavelength range from 1700
 to 6000 \AA.  There 
 was a bright star (B$\sim$12.0 mag) about 10$''$ from RX
 J0148.3-27758 that made it necessary to carry out the UVOT photometry
 using a $4.5\arcsec$ source extraction region. This is smaller than
 the $6\arcsec$ and $12\arcsec$ radii that are used for the optical
 and UV filters, respectively, for the compatibility with the current
 effective area calibrations. An aperture correction was therefore
 applied to account for source photon counts that lay outside of this
 extraction region, in the wings of the PSF. The background region was
 taken from an annulus around the source, off-centered by $\sim7^{''}$
 to avoid excessive contamination from the nearby star. Source photon
 counts, magnitudes and fluxes were then extracted using the UVOT tool
 {\it uvotmaghist} version 1.0 for every individual exposure taken in
 each filter, as well as from the co-added exposures within each
 segment number. All UVOT magnitudes were corrected for Galactic
 reddening with $E_{\rm B-V}$=0.017.

In order to be able to carry out broadband spectral fitting, source
and background data files compatible with XSPEC were created from the
co-added exposures from the 2005 December 09 (segment 009)
observations.  This was done using the tool {\it uvot2pha} version
1.1.  This provided a single spectral file per filter.  The same
source and background extraction regions were used as before, and the
exposure times in the headers were changed to normalize the count
rates to the rates with the aperture correction taken into account.

The field of RX J0148.3--2758 was also observed by the BAT. However, a
preliminary analysis of the BAT pointed and survey data does not show
a detection of the source.  So far more than 100 AGN have been
detected by the BAT, of which about 50 have had the results published
\citep{markwardt06}.

RX J0148.3--2758 was observed by {\it ROSAT} with the Position Sensitive
Proportional Counter \citep[PSPC, ][]{pfeffermann87} three times
during the RASS for a total of 504 s and for 6.7 ks in a pointed
observation (Table \ref{obs_log}).  Source counts were selected in a
circular region with R=200$^{''}$. For the RASS observations
background photons were taken from two circular regions with
R=400$^{''}$ in the {\it ROSAT} scan direction \citep[for details see
][]{belloni94}. For the pointed observation the background was
estimated from a close-by circular region with R=400$^{''}$. Spectra
were rebinned to have at least a S/N=5 in each bin. The light curves
were binned in 400s bins.  The {\it ROSAT} data were processed using the
EXSAS version Apr01 \citep{zim98}.

{\it ASCA} observed RX~J0148.3-2758 on 1997 November 7 for a total of
33.2 ks with its Solid-state Imaging Spectrometers (SIS) and 36.4 ks
with the Gas Image Spectrometers (GIS) on 1997-07-11 (Table
\ref{obs_log}).  A standard configuration was used during the
observation.  The Gas Imaging Spectrometers (GISs) were operated in PH
mode throughout the observation.  The Solid-state Imaging
Spectrometers were operated in 1-CCD Faint mode.  The SIS energy gain
was reprocessed using the latest calibration file ({\sf
sisph2pi\_290301.fits}).  We used standard criteria for reducing the
{\it ASCA} data.  For the SIS detectors, source photons were extracted from
a circular region $3.5^{\prime}$ in radius, and for the GIS detectors,
the source extraction region is $5.25^{\prime}$ in radius.  In both
cases, the background was drawn from source-free regions of the
detectors.

For spectral fitting, the spectra were grouped so that there at least
20 photons pre bin.  It has been demonstrated that the SIS spectra
suffered degradation during the mission.  The SIS efficiency loss can
be parameterized by adding additional absorption to the model, where
the amount of additional absorption depends on the time of the
observation\footnote{see
http://heasarc.gsfc.nasa.gov/docs/asca/calibration/nhparam.html for
details.}.  For the time of the RX~J0148.3-2758, the appropriate
additional column is $4.01 \times 10^{20}\rm \, cm^{-2}$.  We fit the
SIS0 spectrum between 0.5 and 8.0 keV, the SIS1 spectrum between 1 and
0.8 keV, and the two GIS spectra between 0.8 and 8.0 keV.

\section{\label{results} Results}

\subsection{X-rays}

\subsubsection{Light curves}
\begin{figure*}
\epsscale{1.1}
\plottwo{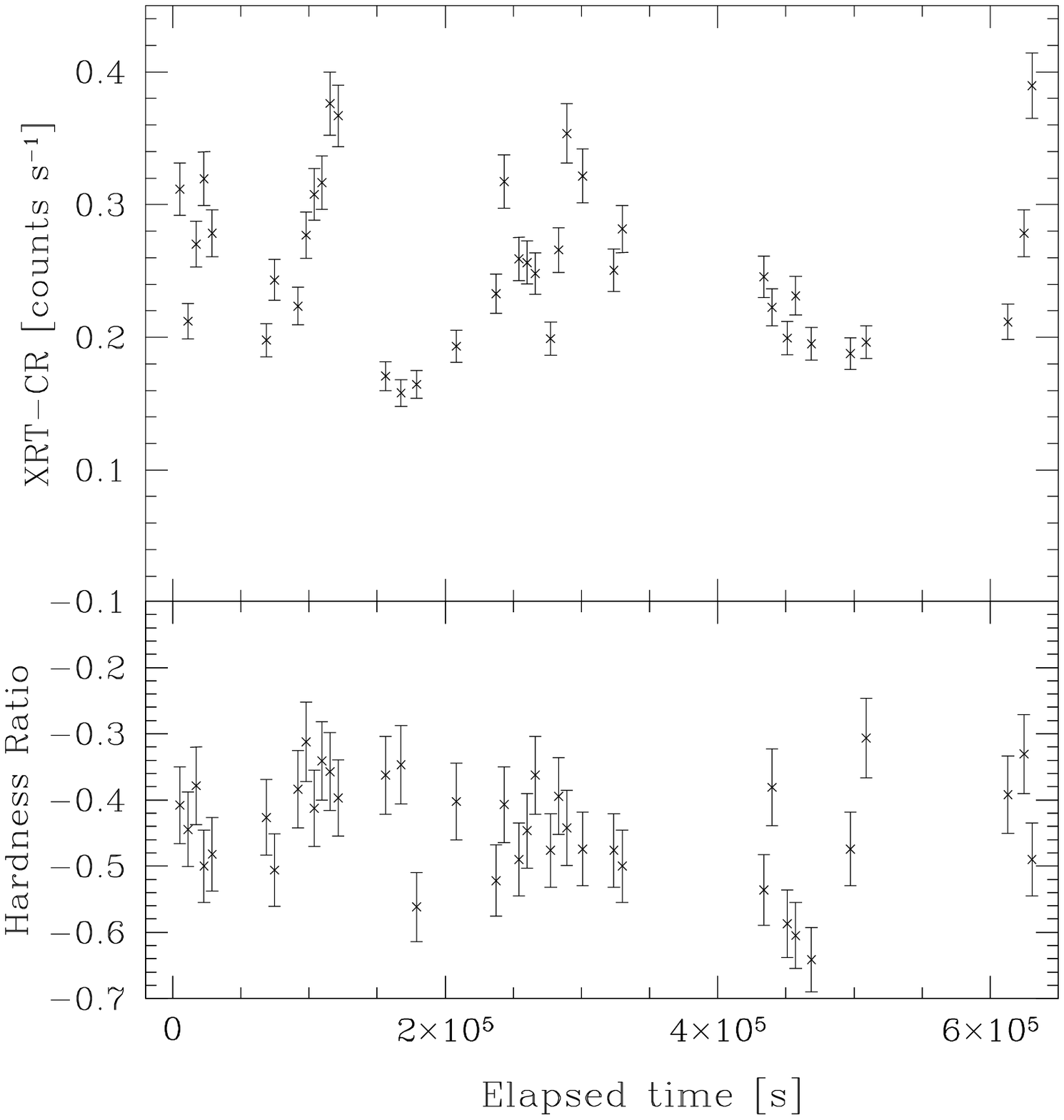}{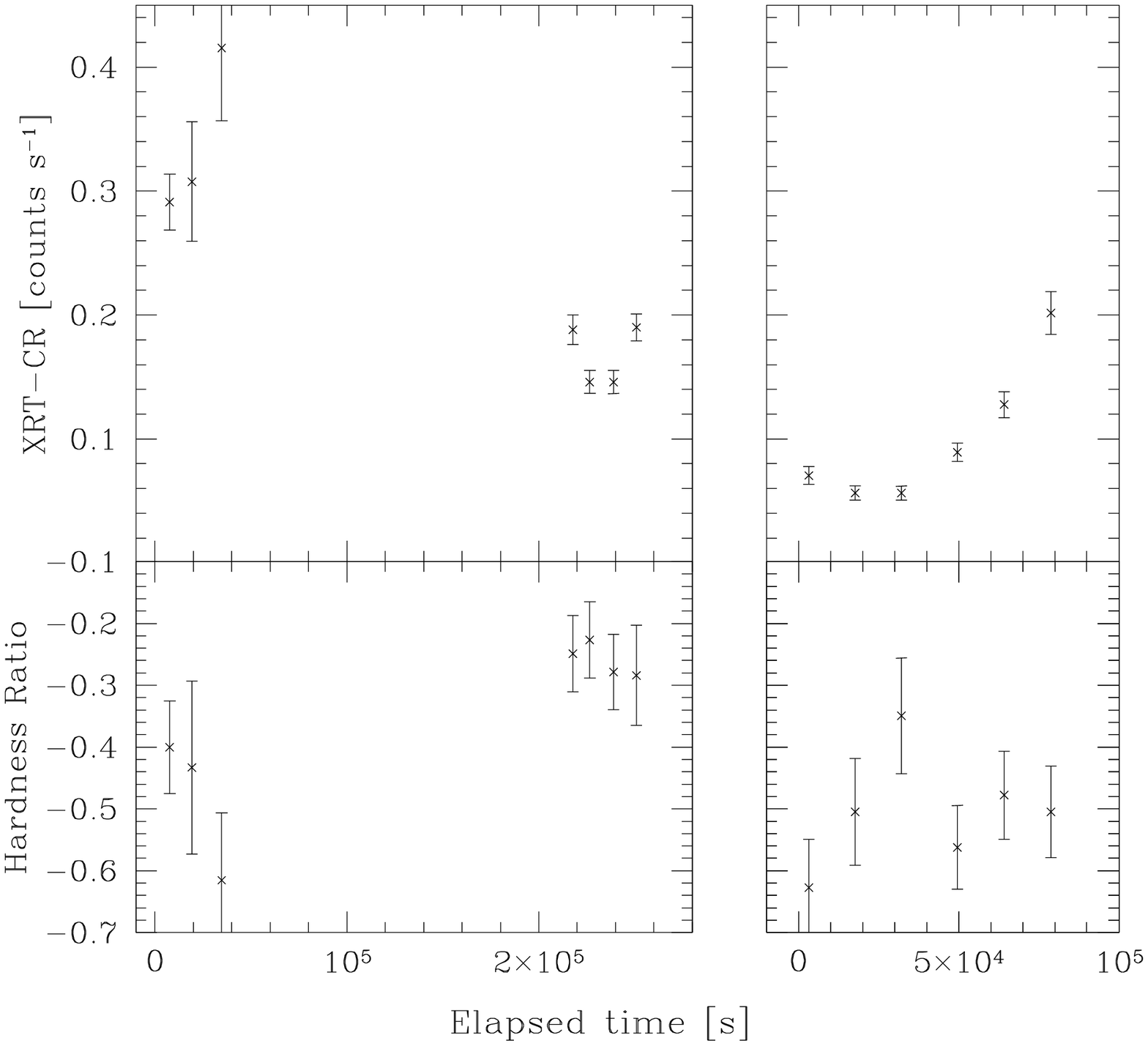}
\caption{\label{rxj0148_lc_swift} \swift~XRT 
0.3-10.0 keV light curves  The left panel shows the 2005 May
observation; the middle panel shows the  December 07 and 09
observations, and the right one shows the December 20 and 21 observations.
Start times are
2005-May-06 00:05 UT, 2005-December-07 00:34 UT, and 2005-December-20 13:40 UT, 
for the left, middle and right panels, respectively.
The start and end times of each segment are given in
Table\,\ref{obs_log}. The upper panel displays the count rate light curve and
the lower panel the light curve of the hardness ratio = (H-S)/(H+S) with S and H 
are the number of photons in in the 0.3-1.0 and 1.0-10.0 keV band, respectively. 
}
\end{figure*}

The left panel of Figure \ref{rxj0148_lc_swift} shows the \swift-XRT
light curve of \rxj~during the 2005 May observations (segments
002-006). The middle and right panels show the observations from 2005
December 07-09 (segments 008+009) and 2005 December 20/21 (segments
010+011), respectively.  The XRT count rate light curves shown in the
upper panels of Figure \ref{rxj0148_lc_swift} suggest that RX
J0148.3-2758 is a highly variable source. In general the AGN varies
between $\approx$ 0.1 to 0.4 count s$^{-1}$.  The most dramatic
variability can be seen in the May 2005 light curve (left panel of
Figure \ref{rxj0148_lc_swift}), where the count rate doubled in 25 ks
between 90--115 ks, followed by a rapid drop between 120--150 ks by a
factor of more than 2. A similar increase in count rate was also
observed at the end of the May 2005 observation when \rxj~ doubled its
count rate within 18 ks. The 2005 December observations show that
during the December 09 observation (middle panel) \rxj~ became
significantly fainter with a count rate of about 0.18 counts
s$^{-1}$. The AGN count rate fell by a further factor of 3 when it was
re-observed on 2005-December-20 (segment 010, right panel). Whether
\rxj~remained in this low state during the 10 day gap . We do not
know what happened during the between 2005 December 09 and
20, whether
\rxj~remained in this low state during the 10 day gap, or whether
further variability took place. By
the end of segment 011 the count rate went back to its `normal' level,
having increased by a factor of about 4 within 30 ks.  The
observations of 2005 December 21st were discontinued at the end of
segment 011 due to the trigger of GRB 051221A \citep{parsons05,
burrows06} which superseded the \rxj~ observation.

The hardness ratio\footnote{The hardness ratio is defined by
(hard-soft)/(hard+soft) with {\it soft} are the count in the 0.3-1.0
keV band and {\it hard} in the 1.0-10.0 keV band.} plots suggest the
presence of some
spectral variability.  While the hardness ratios during the December
07 and 20/21 observations are similar to the ones measured during May
2005, the hardness ratio of the December 09 observations are
significantly harder, suggesting a change in the X-ray spectrum.

\begin{figure*}
\epsscale{1.0}
\plottwo{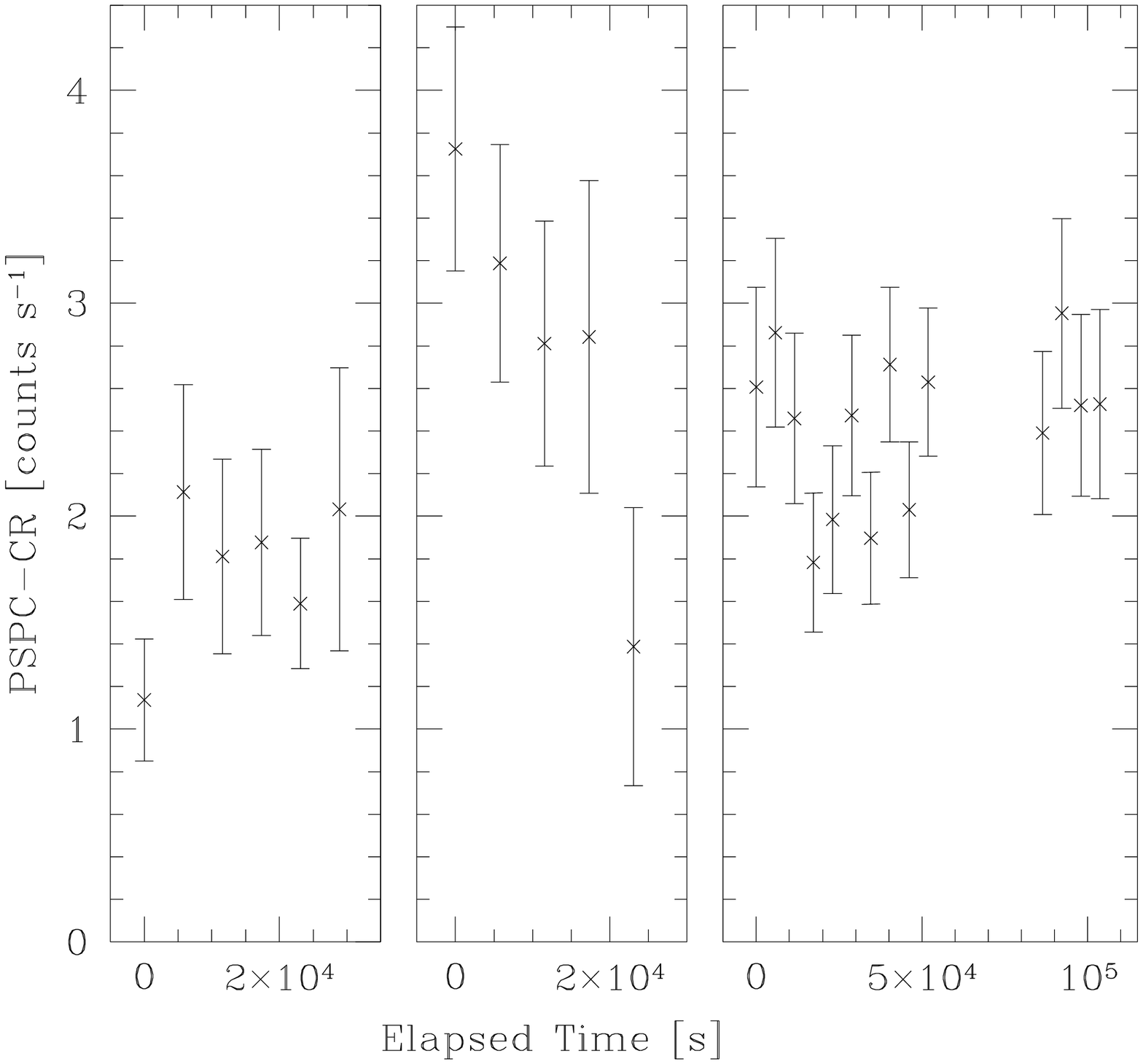}{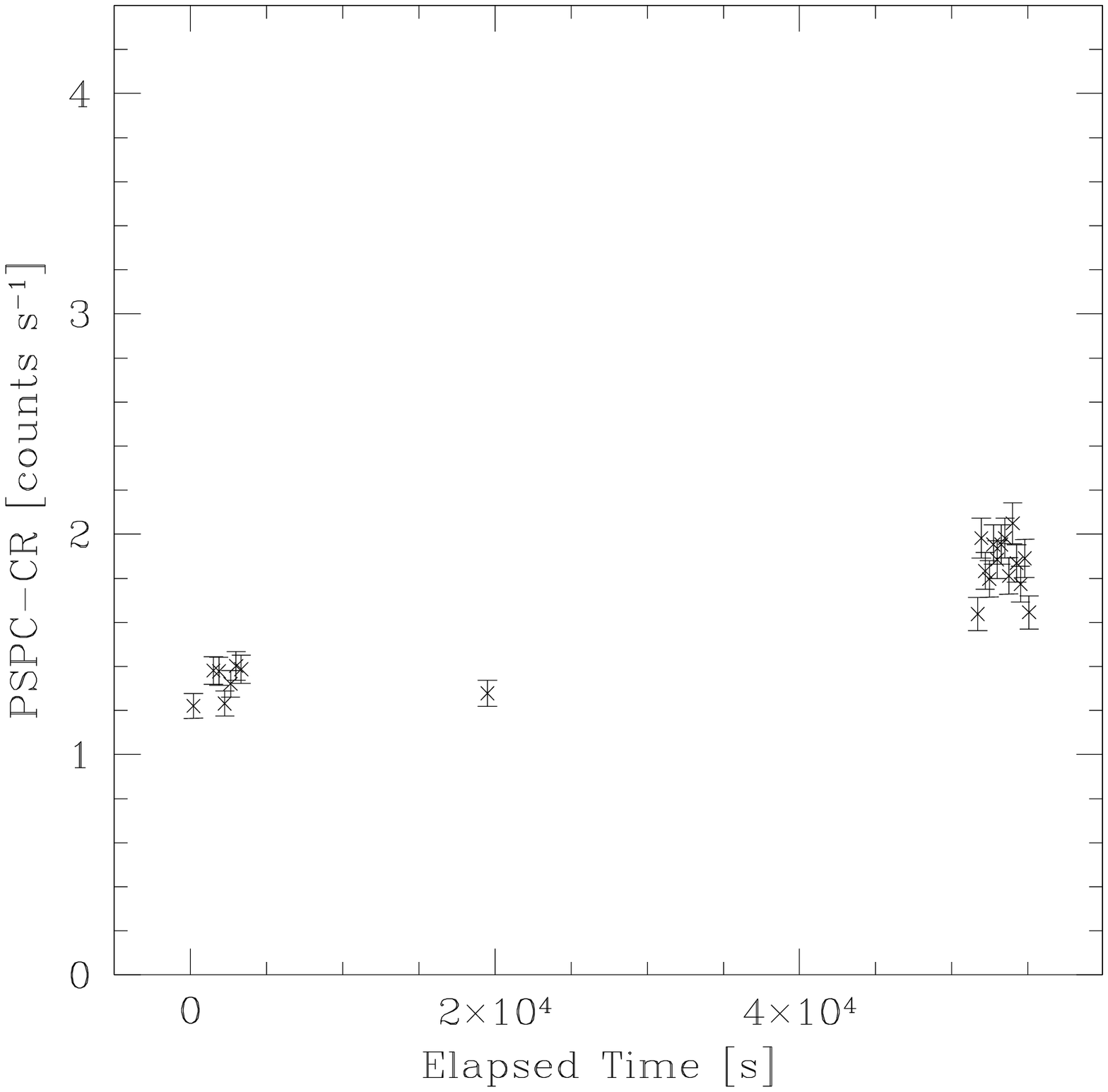}
\caption{\label{rxj0148_lc_rosat} {\it ROSAT} All-Sky Survey and pointed observation
 light curves. The start times are 1990-07-15 15:26 UT, 1990-12-28 01:01 UT, and
 1991-01-15 09:24 UT for the RASS observations (left panels),
 and 1992-07-09 09:54 for the
 pointed {\it ROSAT} PSPC observation (right panel).
  }
\end{figure*}

The left panel of Figure\,\ref{rxj0148_lc_rosat} shows the RASS light
curves and the right panel displays the pointed {\it ROSAT} PSPC light
curve. In both light curves RX J0148.3--2758 displays a similar
variability, in agreement with the \swift-XRT light curve
(Figure\,\ref{rxj0148_lc_swift}).  RX J0148.3--2758 was one of the
most variable of the soft X-ray selected AGN sample of \citet{gru01}.

\begin{figure*}
\epsscale{0.5}
\plotone{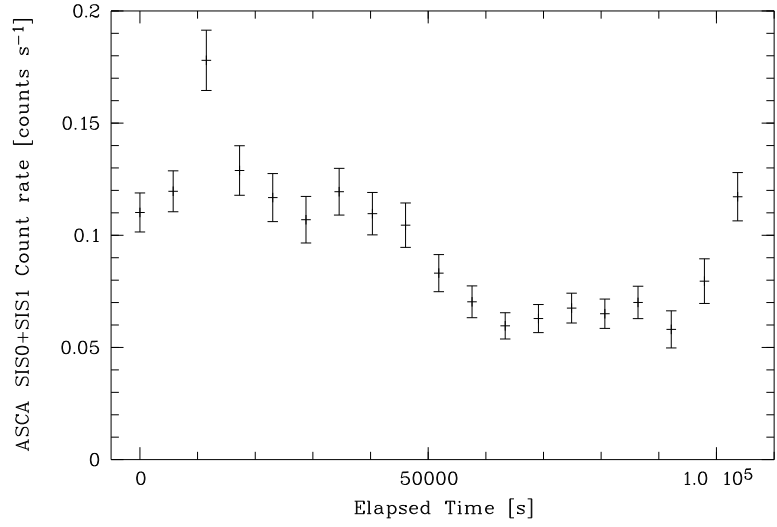}
\caption{\label{rxj0148_lc_asca} {\it ASCA} SIS0 + SIS1 0.5--10 keV
  light curve. The start time was 1997-July-1 21:05 UT.}
\end{figure*}

RX J0148.3--2758 was also observed by {\it ASCA} for a period of about 1
day.  Light curves were extracted in the 0.5--10 keV band for SIS
detectors, and 0.8--10 keV band for the GIS detectors.  The average
net source count rates were 0.054, 0.043, 0.022, and $0.028 \rm\,
counts\, s^{-1}$ in the SIS0, SIS1, GIS2, and GIS3 detectors,
respectively.  The background fraction in the source regions are
estimated to be 20\%, 22\%, 36\% and 30\% in the SIS0, SIS1, GIS2, and
GIS3 detectors, respectively.  The SIS0+SIS1 net count rate light
curve was binned by orbit, and is displayed in Figure
\ref{rxj0148_lc_asca}.  The light curve shows that \rxj~ is varied by
a factor of about 2, consistent with previous findings
\citep[e.g.][]{nandra97}, which found RX J0148.3--2758 to show the
largest excess variance among the AGNs observed by {\it ASCA}
\citep{turner99}.

\begin{figure*}
\epsscale{0.5}
\plotone{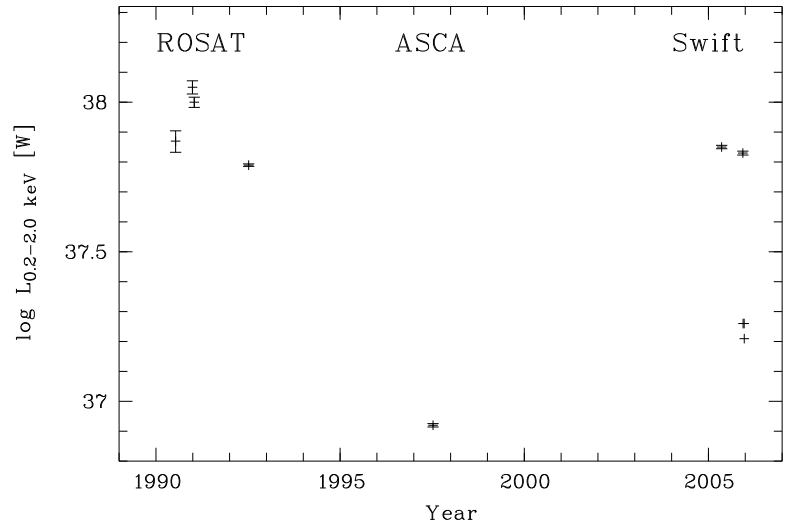}
\caption{\label{rxj0148_lc_all} Long-term light curve of RX J0148.3--2758. 
The luminosities are rest-frame 0.2-2.0 keV and are determined from unabsorbed
fluxes based on the best-fit
models as given in Tables\,\ref{xrt_fit} and \ref{rosat_fit}.
}
\end{figure*}

The long term light curve shown in Figure \ref{rxj0148_lc_all}
displays the complete set of X-ray observations performed on RX
J0148.3--2758. The rest-frame 0.2-2.0 keV luminosities were derived
from the unabsorbed fluxes determined from the best-fit spectral
models as described in \S\,\ref{xray_spec}. All luminosities are
listed in Table\,\ref{obs_log}.  The light curve shows that during the
{\it ASCA} observation RX J0148.3--2758 was in a low state - about 13 times
fainter than during the RASS observation in December 1990. During the
\swift\ observation in May 2005 and at the beginning of the 2005
December observations, \rxj\ was at a similar brightness as in  first
RASS and the 1992 pointed {\it ROSAT} observations.  By the end of the 2005
December \swift\ observations, \rxj\ had become significantly fainter,
only a factor of about 2 brighter than during the {\it ASCA} low-state.

\subsubsection{\label{xray_spec} Spectral Analysis}
\begin{figure*}
\epsscale{0.9}
\plottwo{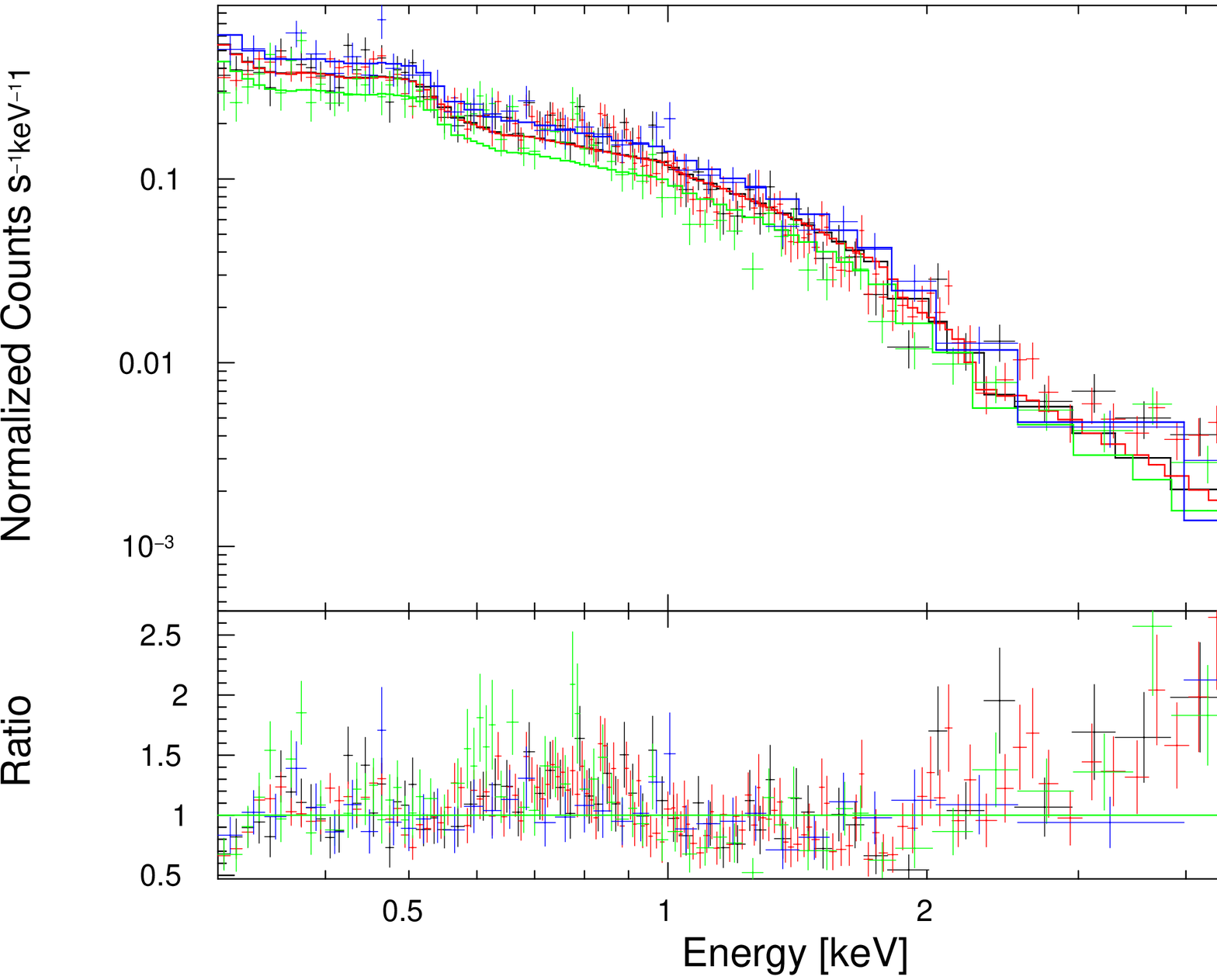}{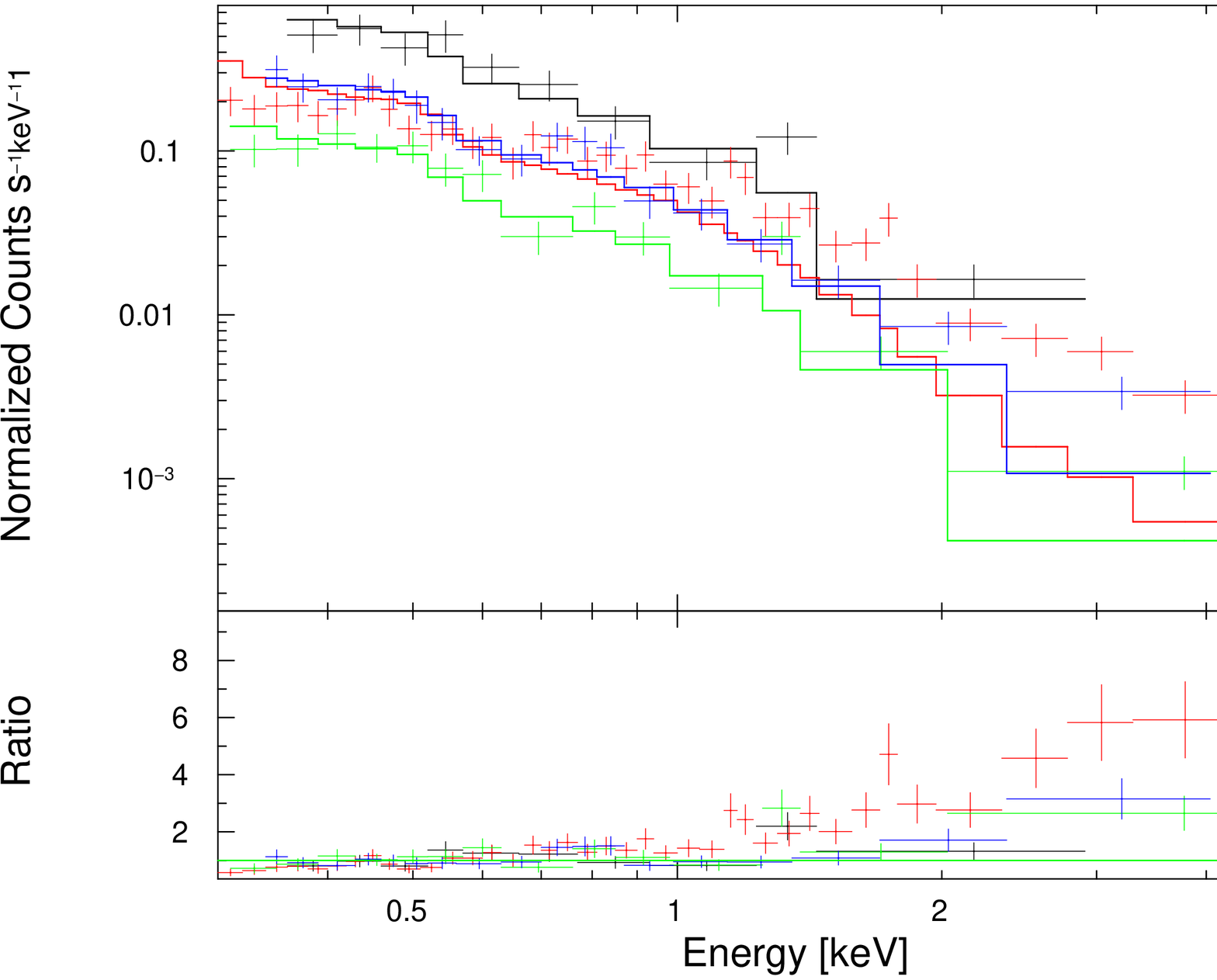}
\caption{\label{rxj0148_xspec} \swift~XRT spectra  of RX J0148.3--2758.
 The 2005 May spectra (left panel) were fitted by a
single power low with the absorption column fixed to the Galactic value.
For the spectra of the 2005 December observations (right panel), 
the X-ray spectral slope was also
 fixed to \ax=2.4 (see text). The colors represent the spectra of different segments: in
the left panel: segment 002 = black, segment 003 = red, segment 004 = green, and
segment 006 = blue; right panel: segment 008 = black, segment 009 = red,
segment 010 = green, and segment 011 = blue.
}
\end{figure*}

Figure\,\ref{rxj0148_xspec} displays the spectra of the 2005-May (left
panel) and the 2005-December (right panel) observations. All spectra
were initially fitted using a single absorbed power law with the
absorption column density at z=0 fixed to the Galactic value
\citep[1.50$\times~10^{20}$ cm$^{-2}$][]{dic90}.  A simple absorbed
power law did not produce 
acceptable fits (Table \ref{xrt_fit}). With the exception of the 2005
December 09 observations, all spectra require multi-component spectral
models such as blackbody plus power law or a broken power law model,
with an additional absorption component above the Galactic column
density. A broken power law model as well as a blackbody plus power
law model yield similar $\chi^2/\nu$ and we cannot distinguish between
the models.  Using a blackbody model over a hard power law component
yields a temperature kT$\approx$100-120 eV which is typical for a NLS1
and agrees with the value kT=120 eV found by {\it ASCA} (Table
\ref{rosat_fit}).  A broken power law model simultaneously fitted to
the 2005 May XRT spectra results in a soft X-ray spectral slope
$\alpha_{\rm X, soft}$=2.58$^{+0.15}_{-0.12}$ which is in good
agreement with the results found by {\it ROSAT} (both the RASS and pointed
observations; Table \ref{rosat_fit}).  The XRT hard X-ray spectral
slope $\alpha_{\rm X, hard}$=0.96$^{+0.16}_{-0.14}$ is also in good
agreement with the hard X-ray spectral slope from the {\it ASCA} data (Table
\ref{rosat_fit}). In addition to the phenomenological models we fitted
the partial covering absorber model {\it pcfabs}, the `warm', ionized
absorber model {\it absori}, the reflection model {\it pexrav}, and
the disk blackbody model to the 2005 May spectra.  We found the {\it
absori} and {\it disk blackbody} models to show no improvement over a
simple powerlaw model, and the parameters of the reflection model {\it
pexrav} could not be constrained. However, the partial covering model
{\it pcfabs} yields reasonable results. We found the 2005 May spectra
to be well fit by a partial covering absorber with a column density
$N_{\rm H}$=7.2$^{+1.7}_{-1.3}\times 10^{22}$ cm$^{-2}$, a covering
fraction $f=0.80^{+0.03}_{-0.04}$, and a spectral index
\ax=2.34\plm0.09 where $\chi^2/\nu$=428/302, which is significantly
better than a single power law fit as listed in Table\,\ref{xrt_fit}.

Although the 2005 December 07 observations provided poorly constrained
spectral fits due to the small number of photons (290 in 822 s), some
interesting spectral variability is observed in the 2005 December data
set. The hardness ratio light curve (lower right panel of Figure
\ref{rxj0148_lc_swift}) suggests that \rxj~had a similar spectrum as
during the 2005 May and 2005 December 20/21 observations.  However, a
fit to the 6.3 ks observation from 2005 December 09 with a single
absorber at z=0 yields an absorption column density consistent with
the Galactic value, suggesting there is no intrinsic
absorption. Furthermore, an absorbed broken power law model yields a
soft X-ray spectral slope \axs=1.93$^{+0.58}_{-0.42}$ flatter than
during the 2005-May observations. To examine this spectral change, a
Target-of-Opportunity observation was made with \swift\  on 2005 December
20/21. The spectral analysis of these data show \rxj~to have once
again become intrinsically absorbed, with $N_{\rm H,
intr}=11.6^{+7.5}_{-5.7}\times 10^{20}$ cm$^{-2}$, and the soft X-ray
spectral slope \axs=3.41$^{+0.78}_{-0.64}$ to have become
significantly steeper. It is also interesting to note that between the
2005 May and December observations the best-fit spectral break from a
broken power law model fit has shifted towards softer energies. During
the 2005 May observations the break energy was at $E_{\rm
break}=1.68^{+0.12}_{-0.14}$ keV while during the 2005 December
observations the break energy shifted to $E_{\rm break}\approx 1.2$
keV in agreement with the break energy $E_{\rm
break}$=1.36$^{+0.16}_{-0.19}$ found by {\it ASCA} when \rxj~was also in a
low state.  As a result of the low number of photons in the December
09 observation the parameters of a partial covering absorber were
poorly constrained if all parameters were left to vary. However, by
fixing the spectral index to the value during the 2005 May
observations, \ax=2.33, the fit resulted in a partial covering
absorber with essentially the same covering fraction
$f$=0.75$^{+0.06}_{-0.09}$ as during the 2005 May observation, but
with a significantly lower \nh=$1.9^{+1.7}_{-0.9}\times 10^{22}$
cm$^{-2}$ with $\chi^2/\nu$=35/35.  A partial covering absorber model
fit to the December 20/21 spectrum find an increase of the absorption
column of the partial covering absorber to
\nh=3.6$^{+3.2}_{-1.4}\times 10^{22}$ cm$^{-2}$, a covering fraction
$f=0.87^{+0.06}_{-0.12}$ and \ax=3.98$^{+0.48}_{-0.44}$, with
$\chi^2/\nu$=30/27. 

The left panel of Figure \ref{rxj0148_contour} displays the spectra
from the merged data sets of the 2005 May observations, the data set
from 2005 December 09, and the merged data set from December
20/21. The right panel shows the corresponding contour plots between
the intrinsic column density and the photon index $\Gamma =
\alpha_{\rm X}+1$.  The lack of overlaps between the 2005-May, 2005 
December 09, and 2005 December 20/21 observations suggest the presence
of significant spectral variability in \rxj.  The spectra in the left
panel of Figure \ref{rxj0148_contour} show how the spectra change:
compared with the 2005 May observations, the spectrum from December 09
has a similar hard X-ray flux, but a significantly lower flux in the
soft X-ray component.  Then in the December 20/21 observations the
soft X-ray component remained at a similar level as the December 09
observation. The hard X-ray flux, however, decreased by a factor of
4. By the end of the December 21st observation \rxj~ increased its
0.3-10.0 observed flux by a factor of 3 (Figure
\ref{rxj0148_lc_swift}).

The short-term variability we observed in \rxj~during the \swift\
observations seems to reflect the previous measurements by {\it ROSAT} and
{\it ASCA}. During the high state observations during the RASS and {\it ROSAT}
pointed observations the soft X-ray spectral slope was steep with
\axs=2.62 and 2.25, respectively.  For the spectral analysis of the
{\it ASCA} data, we first constrain the power law index by fitting the
region between 2 and 5 keV with a power law model.  We obtain a good
fit ($\chi^2=158$ for 173 degrees of freedom) and measure the energy
index to be $1.16^{+0.27}_{-0.26}$.  Next, we include the photons
between 5 and 8 keV.  The residuals show a slight excess that may
indicate the presence of a reprocessing component.  Indeed, when we
plot the spectrum in this bandpass, we find the photon index flattens
to 2.01, although the difference is not significant.  We add a narrow
iron line at 6.4 keV, but find no significant decrease in $\chi^2$
($\Delta\chi^2=1.78$). Allowing the line energy to vary yields a better
fit with $6.70^{+0.18}_{-0.24} \rm\, keV$, and a greater reduction in
$\chi^2$. However, the line equivalent width is very large (550 eV),
and an F-test indicates the change in $\chi^2$ of 5.9 compared with the
no-line model to not be a significant improvement. Similarly, a broad
line does not improve the fit significantly, and produces a line with
unphysically large equivalent width.  We conclude that evidence for a
line in these data is weak, most likely because of the low
signal-to-noise ratio at high energies in the spectrum.  Since there
is some flattening that distorts the powerlaw, we ignore the spectra
above 5 keV henceforth.  Note that due to the lower effective area in
the \swift~XRT at 6 keV we were not able to identify the line with the
XRT.

Next, we examine the spectrum at low energies.  Extrapolating down to
the lower limits described above, we find that the continuum subtly
steepens toward low energies.  Indeed, fitting between 1 and 5 keV
gives an energy index of $1.25 \pm 0.10$, while fitting down to the
lowest limits on the spectrum yields $1.48^{+0.09}_{-0.08}$.  We
conclude that there is a weak soft excess present.  We can model the
soft excess with either a blackbody or broken power law. The fit
parameters between 0.5 and 5 keV are given in Table \ref{rosat_fit}.
The soft X-ray spectral slope \axs=2.03$^{+0.23}_{-0.20}$ is in good
agreement \axs=1.93$^{+0.58}_{-0.42}$ found during the 2005 December
09 observation by \swift.

\subsection{UVOT Photometry}
\begin{figure*}
\epsscale{0.9}
\plotone{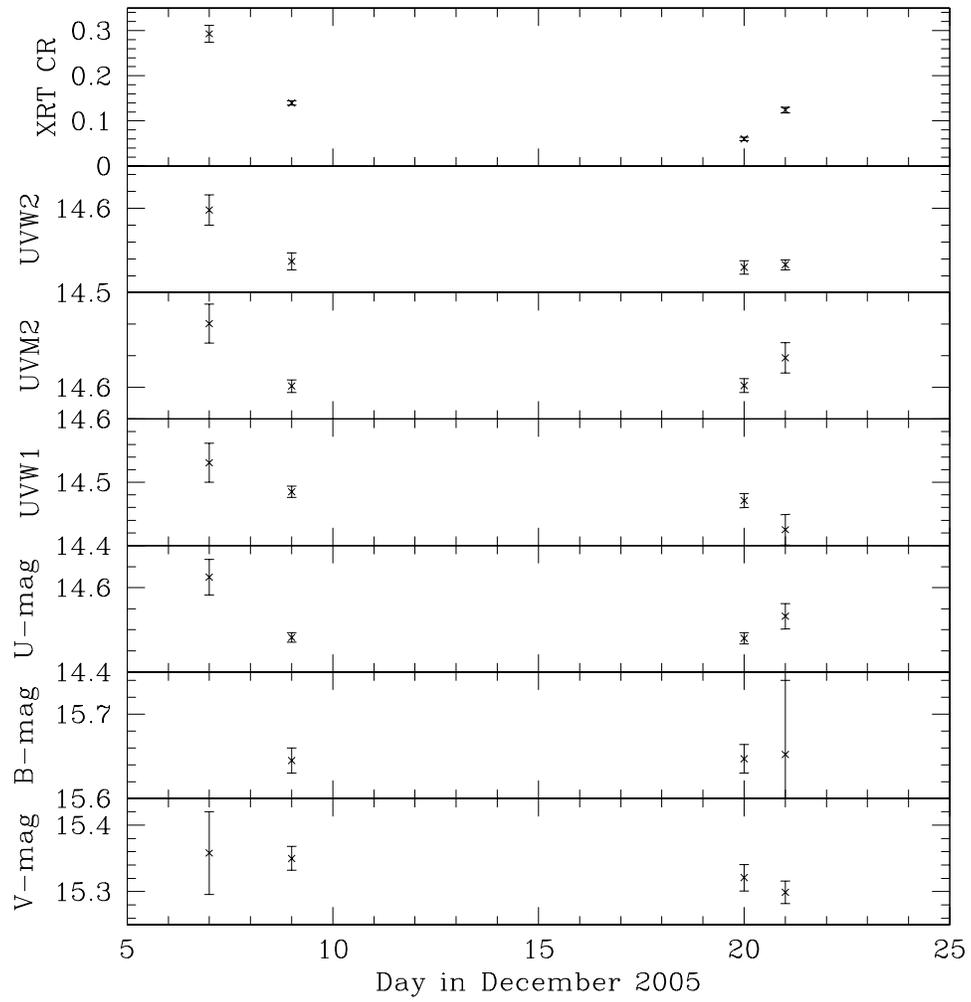}
\caption{\label{rxj0148_uvot} XRT and UVOT light curves. }
\end{figure*}

\begin{figure*}
\epsscale{0.9}
\plottwo{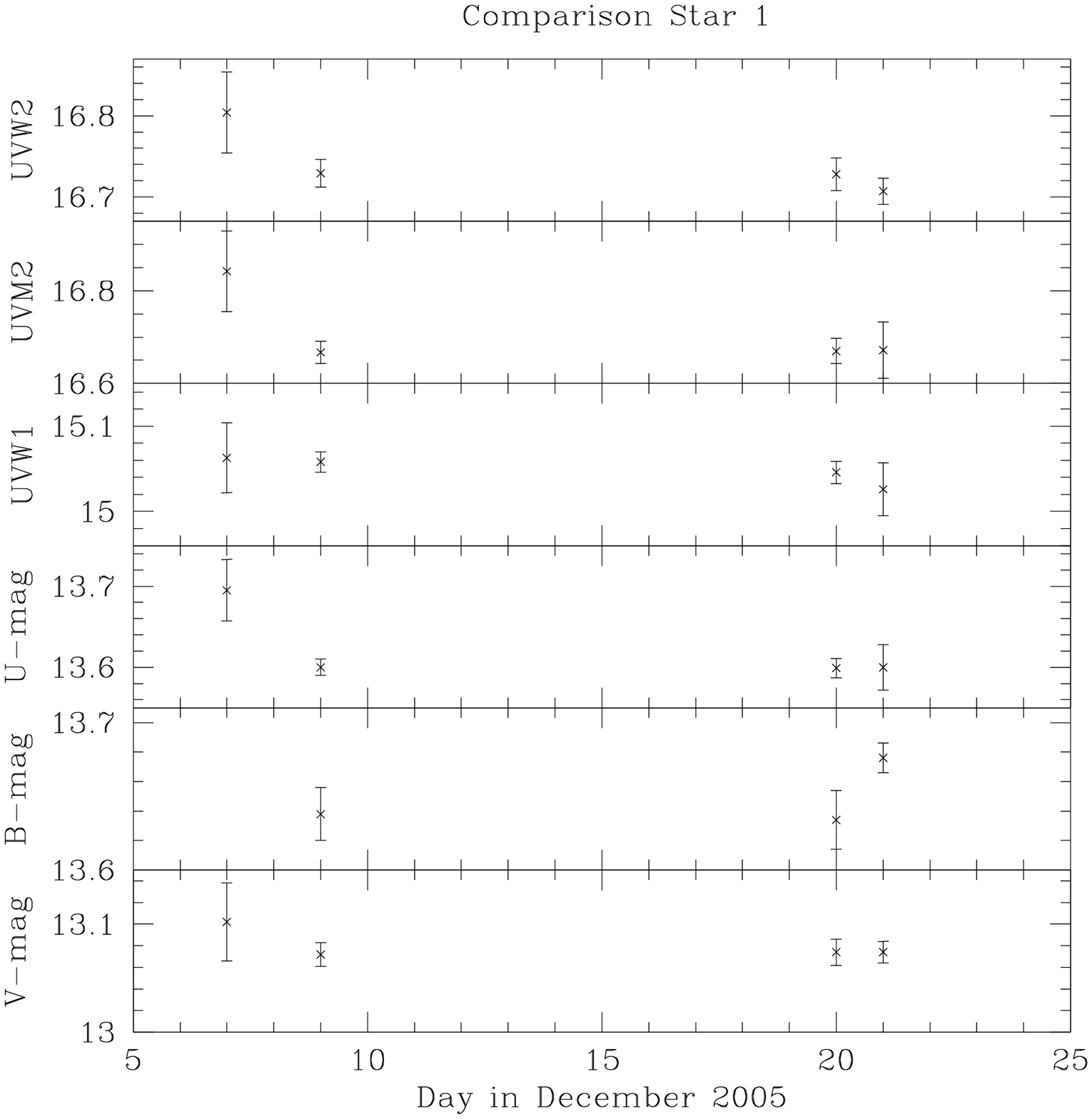}{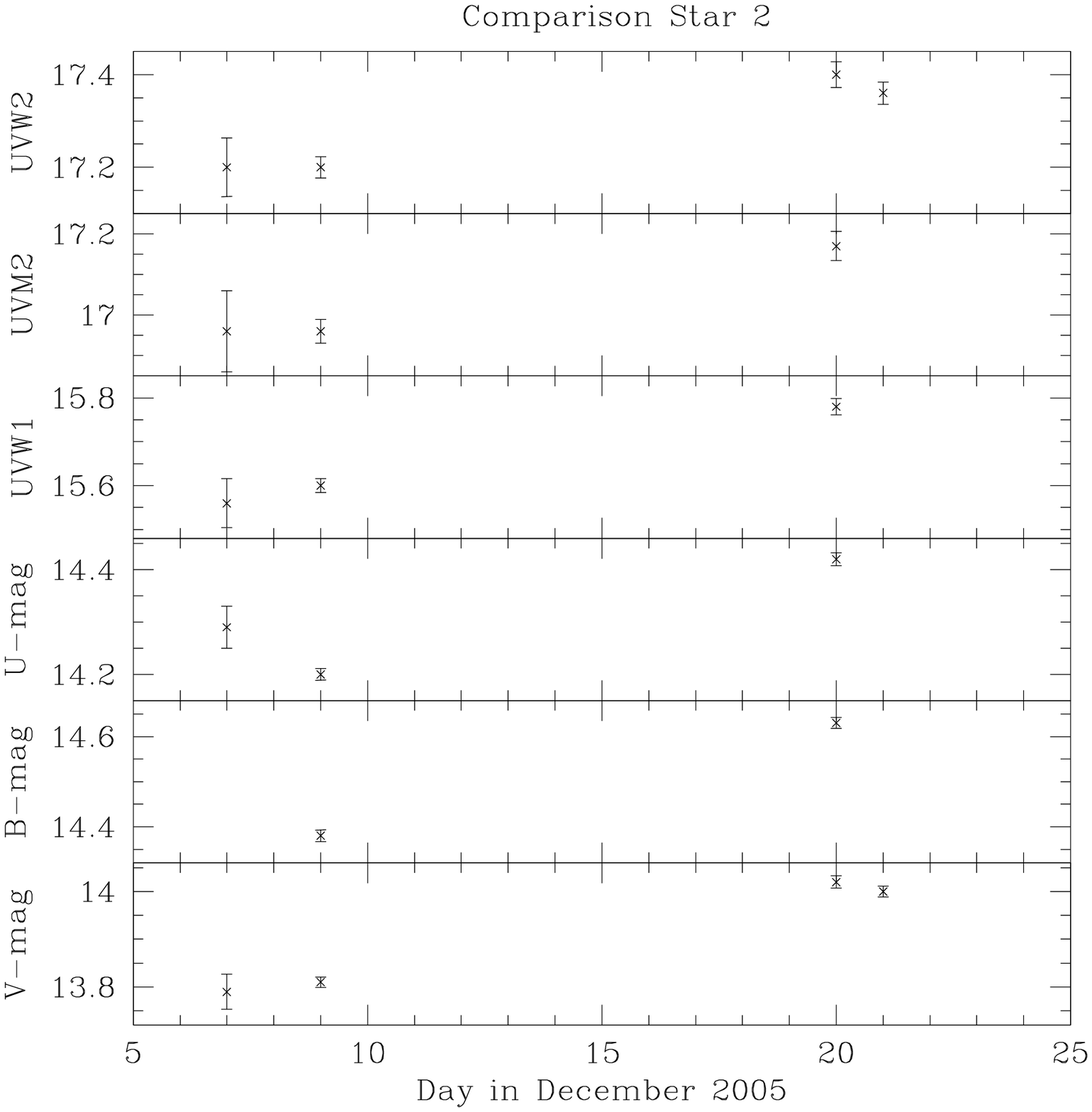}
\plottwo{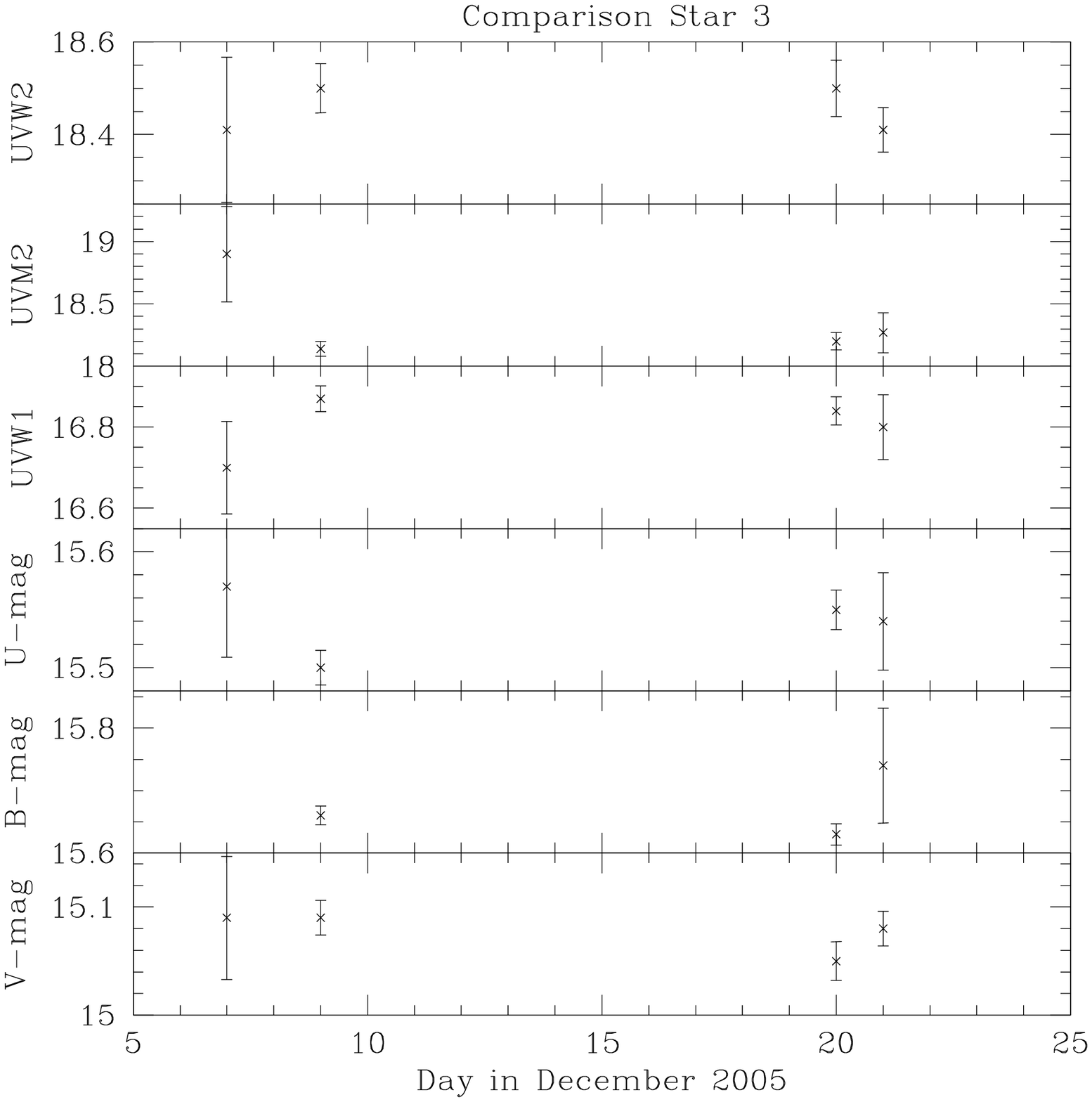}{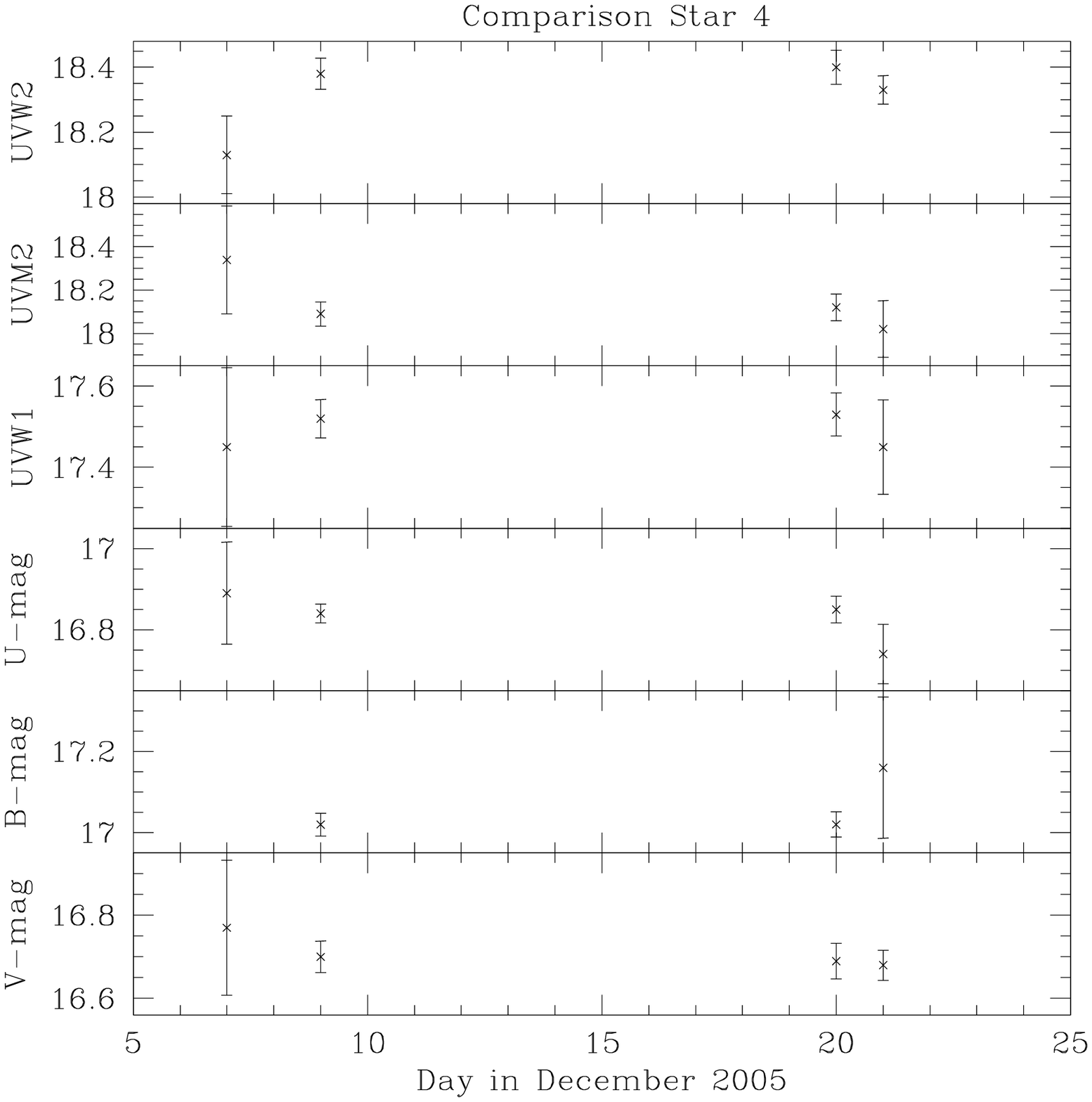}
\caption{\label{rxj0148_uvot_stars} UVOT light curves of the comparison stars S1-S4 as
given in Table\,\ref{stars} and displayed in Figure\,\ref{rxj0148_image}. }
\end{figure*}

\begin{figure*}
\epsscale{0.8}
\plotone{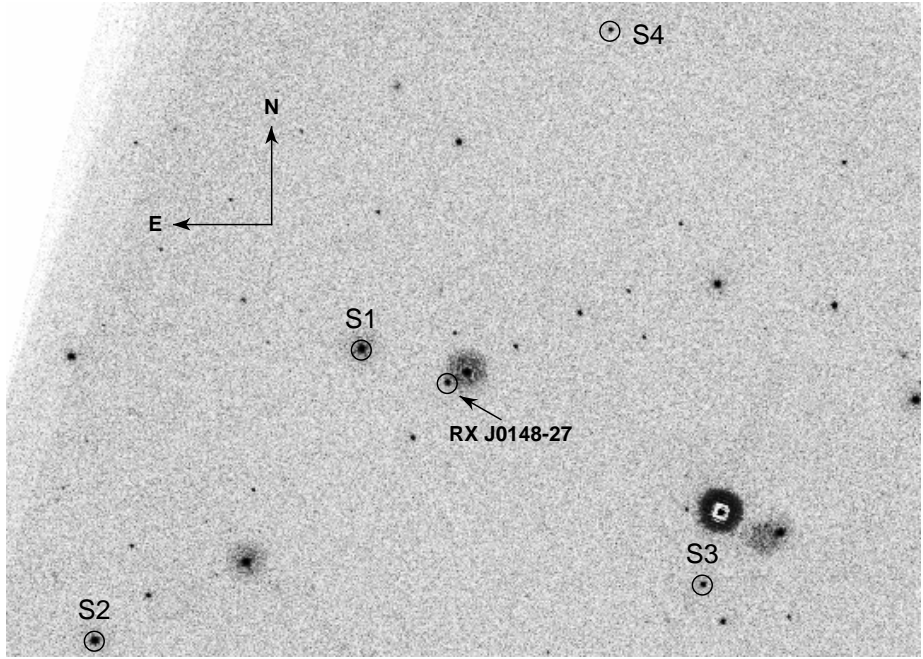}
\caption{\label{rxj0148_image} UVOT V-image of the field around \rxj. The 4
comparison stars as listed in Table \ref{stars} are marked as S1 - S4 in the figure. }
\end{figure*}

Table \ref{uvot_photometry} summarizes the results of the analysis of
the photometry of the co-added UVOT images. During segment 008, no
observations of \rxj~were made in the B filter.  Figure
\ref{rxj0148_uvot} displays the UVOT light curves of all 6 filters
plus the XRT light curve from segments 008 to 011. The figure might
suggest that there is some variability in the UV. However a comparison
with 4 field stars, as listed in Table \ref{stars}, shows that the
variation seen in the UVOT light curves are still within the error
margins.  Figure\,\ref{rxj0148_uvot_stars} displays the UVOT
measurements of these comparison stars. This figure shows that the
trends seen in the \rxj~UVOT light curves are also present in the
light curves of the comparison stars.  Therefore we consider \rxj\ not
to be  variable in the UV/optical band during 2005 December observations.
Figure \ref{rxj0148_image} displays the UVOT V image of the field
around \rxj~with the 4 comparison stars marked.  In part, the
variations seen in the UV light curves of \rxj~ are due to the
relatively small extraction radius of 4.5$^{''}$ and the variable PSF
of the UVOT.  However, the
lack of photometric data during the 2005 May observations provide us
with no knowledge on the UV flux/magnitudes during a high-state.

\subsection{Spectral Energy Distribution}

\begin{figure*}
\epsscale{0.8}
\plotone{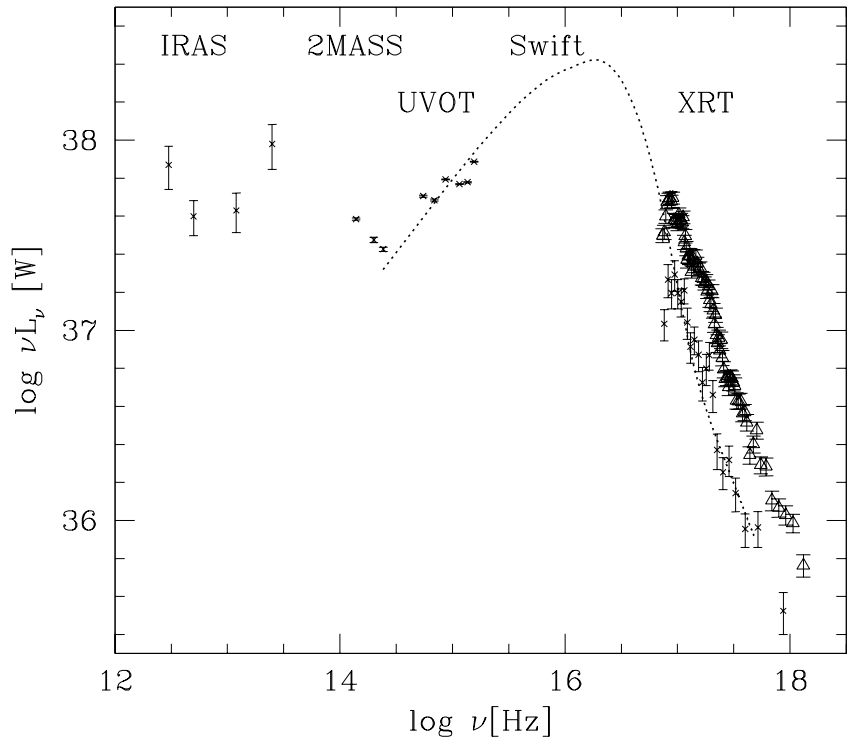}
\caption{\label{rxj0148_sed} Spectral Energy Distribution of RX J0148.3--2758. 
The luminosities are the observed luminosities as given in
Table\,\ref{sed_data}.
The crosses of the \swift~UVOT and XRT data represent the 2005 December
20/21 observations and the triangles the XRT observations from 2005 May. The
dotted line displays the power law plus exponential cutoff and absorbed power
law model to describe the BBB.
}
\end{figure*}

Figure \ref{rxj0148_sed} displays the Spectral Energy Distribution
(SED) of RX J0148.3--2758.  The \swift~XRT data taken in 2005 May are
shown as triangles and the 2005 December 20/21 observations are
represented by diagonal crosses.  For the UVOT data, only the 2005
December data are shown.  This AGN was not detected in the NVSS or the
FIRST radio catalogues. The Far-Infrared {\it IRAS} and NIR 2MASS
luminosities were derived with the {\it GATOR} catalogue search engine
at NASA/IPAC (irsa.ipac.caltech.edu/applications/Gator/).  The {\it IRAS}
luminosities deviate slightly from those given in \citet{gru98a} due
to the improved extraction software at IPAC.

We measured the optical-to-X-ray spectral slopes \aox\footnote{The
X-ray loudness is defined by \citet{tananbaum79} as \aox=--0.384
log($f_{\rm 2keV}/f_{2500\AA}$).} of the 2005 December 09 and 20/21
observations from the SED plot Figure \ref{rxj0148_sed}. During the
December 09 observation we found a rest-frame \aox=1.53.  At a
luminosity density log $l_o$=22.78 [W Hz$^{-1}$] and redshift z=0.121
this value is in good agreement with the mean of radio-quiet AGN {\it ROSAT}
sample of \citet{yuan98a, yuan98b} and \citet{strateva05} for the same
redshift and luminosity intervals. Following the relation given in
equation (4) in \citet{strateva05}, we would expect \aox=1.42.
However, during the December 20/21 observation the source became more
X-ray quiet with an \aox=1.81.  We are yet to analyze the grism data
taken during May 2005, during which the source was in a high state. We
therefore have no current measure of \aox~ during this time
period. Although the 2005 December UVOT observations show no
significant variability, the value of \aox~during the observations of
2005 May cannot be determined because we do not know what the flux in
the UV filters was during this time period.

\begin{figure*}
\epsscale{0.75}
\plottwo{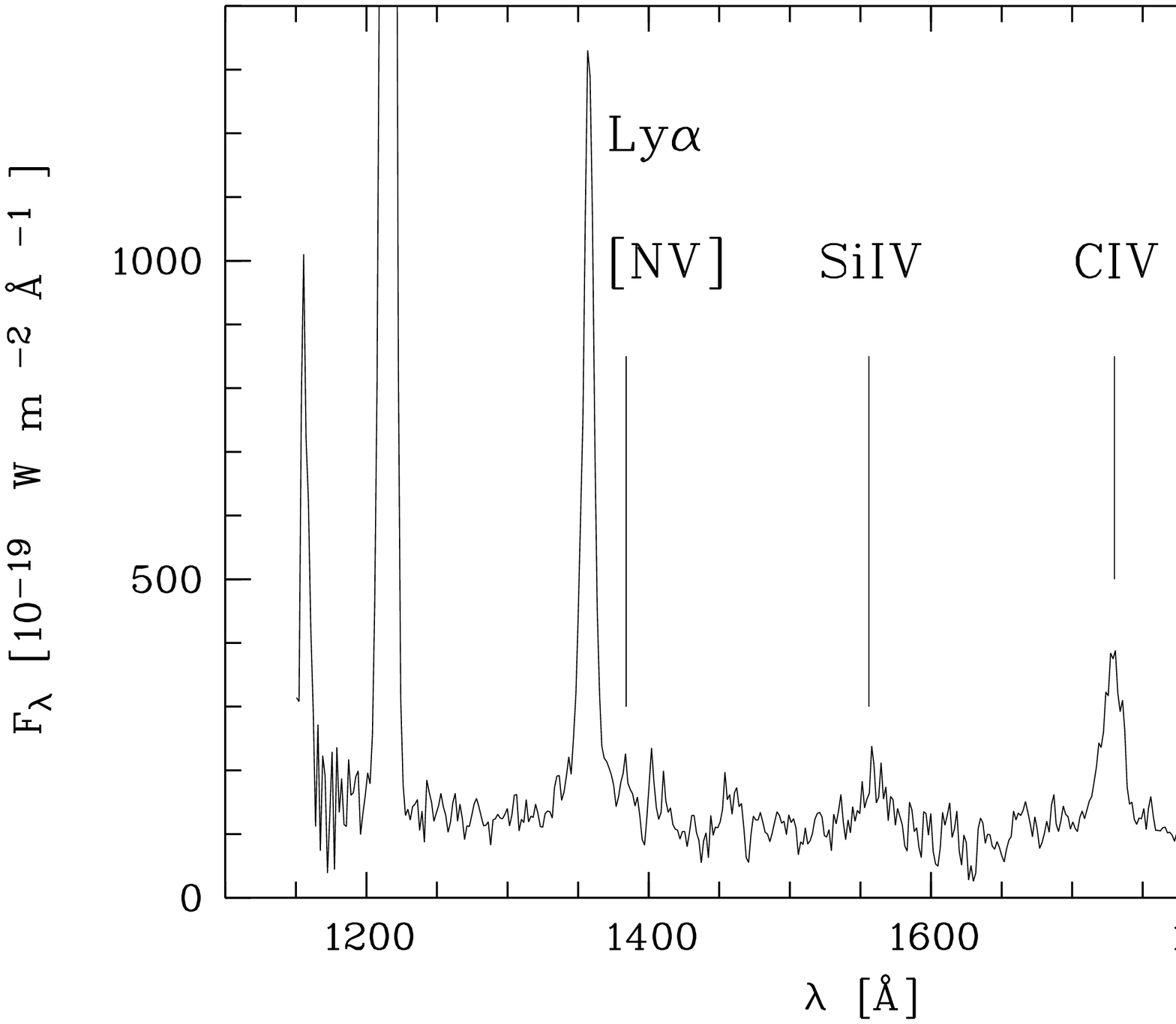}{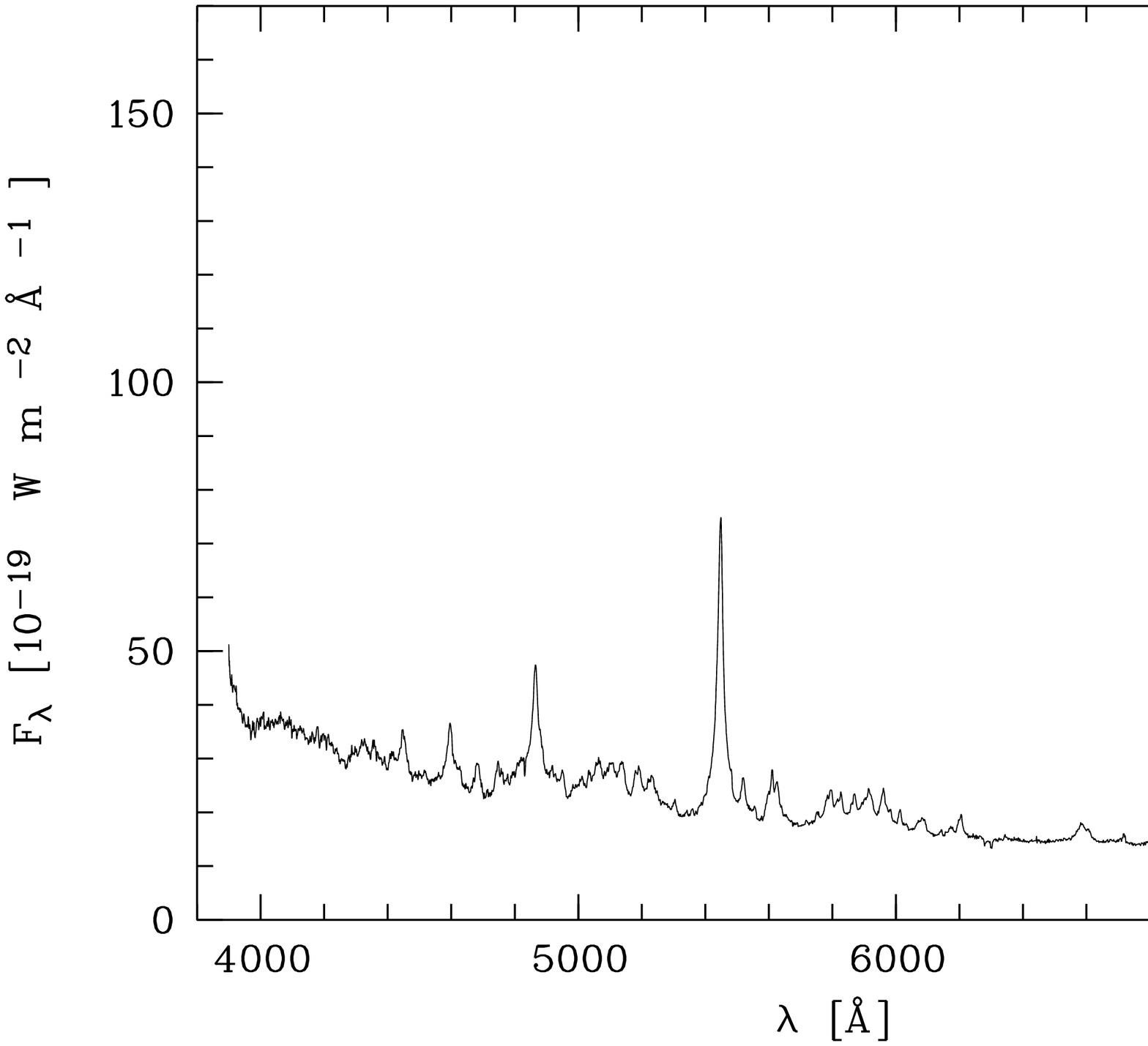}
\caption{\label{opt_spec} IUE and optical spectra of RX J0148.3--2758 }
\end{figure*}

This NLS1 has been observed once before in the UV, in 1992 by IUE (SWP
45107).  The spectrum is displayed in the left panel of Figure
\ref{opt_spec}.  The right panel of Figure \ref{opt_spec} shows the
optical spectrum of \rxj~taken in September 1995 at the ESO 1.52m
telescope in La Silla for a total of 4 hours.  Details of this
observing run are given in \citet{gru04a}. With a
FWHM(H$\beta$)=1030\plm100 \kms~we derived a central black hole mass
of 1.3$\times 10^7$\msun using equation (5) in \citet{vester06}. From
the IUE spectrum shown in Figure \ref{opt_spec} we derived a FWHM(CIV)
= 2300 \kms. By using the relation given in equation (7) in
\citet{vester06} we estimated the black hole mass $M_{\rm
BH}=3.4\times 10^7$\msun. Both black hole masses estimates agree with
each other within their uncertainties.

Estimated from these black hole masses, the Eddington luminosity is
1.6-4.3$\times 10^{38}$ W. As described in \citet{gru04a} we modeled the BBB by
a powerlaw with exponential cutoff plus an absorbed power law. This model is
displayed in Figure\,\ref{rxj0148_sed}. From the 2005 December 20/21 data we
measured a bolometric luminosity $L_{\rm bol}$ = 5$\times 10^{38}$ W which is
similar to the value given by \citet{gru04a} based on the optical spectrum and
the ROSAT RASS data. This results in an Eddington ration $L/L_{\rm Edd}$ = 1-3.

The [OIII] lines can be separated into a narrow and a blueshifted
broad component.  The broad [OIII] lines are blueshifted by 600\plm200
\kms~with respect to the narrow [OIII] and H$\beta$ lines. Similar
results on NLS1s have been previously reported by e.g. \citet{grupe02,
zamanov02, aoki05} and \citet{bian05}.

\section{\label{discuss} Discussion}

In this paper we have presented the \swift\ observations of the high
variable NLS1 \rxj. In addition to the strong X-ray flux variability,
our main results are composed of the spectral changes. We observed a
hardening followed by a softening of the spectrum of \rxj~over a time
during which the X-ray flux was on a continual decrease. Both types of
spectral changes have been observed in AGN, although the hardening
of the X-ray spectrum with decreasing flux is more common
\citep[e.g. ][]{gallo04a, dewangan02, lee01, chiang00}. However, 
softening of the X-ray spectrum with decreasing flux has been reported
on NLS1s in e.g., RX 
J2217.9--5941 \citep{gru04b}, RX J0134.2--4258 \citep{gru00, kom00},
PKS 0558--504 \citep{gliozzi01}, and 1H 0707--495 \citep{gallo04b,
fabian04}.

\begin{figure*}
\epsscale{0.9}
\plottwo{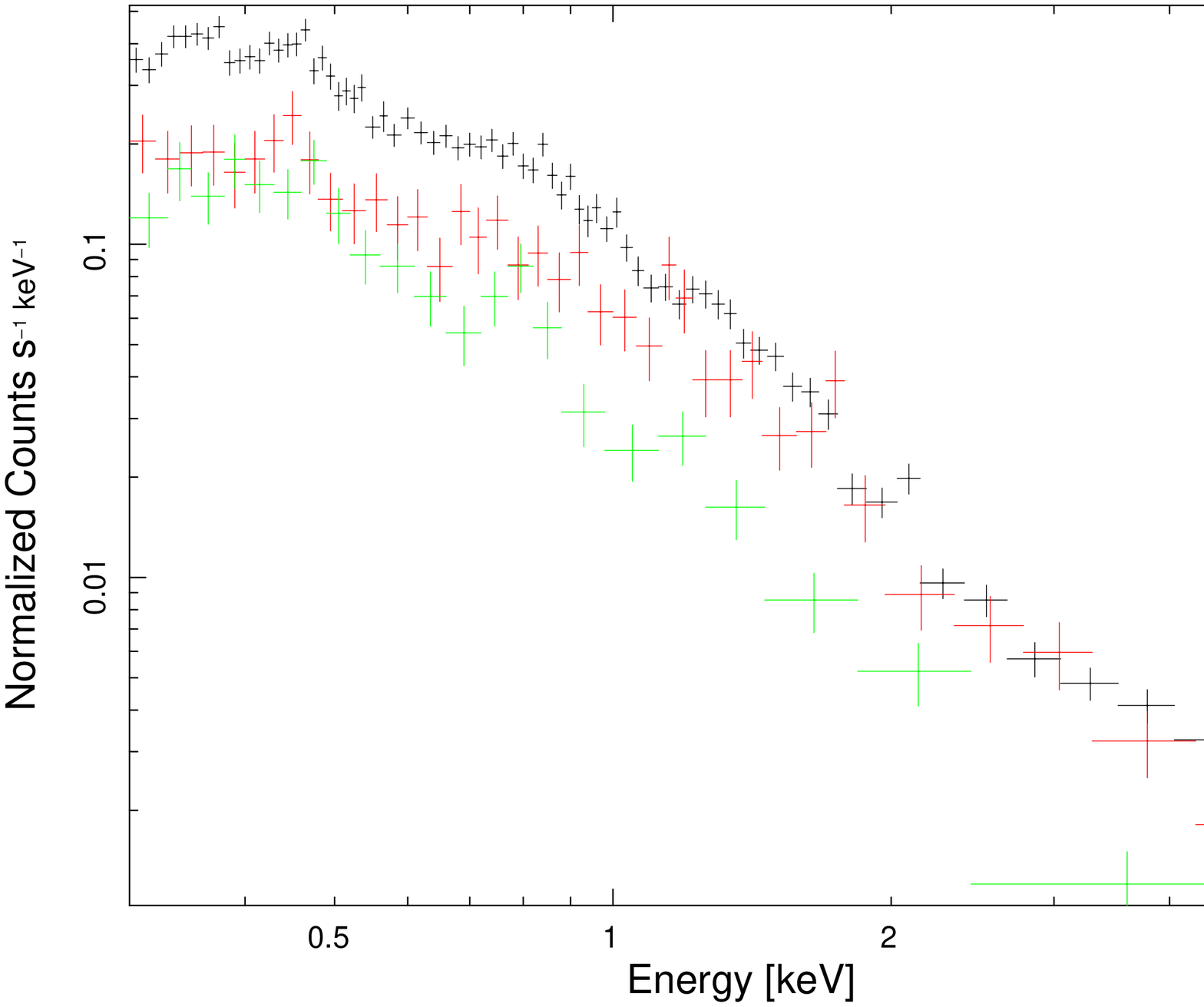}{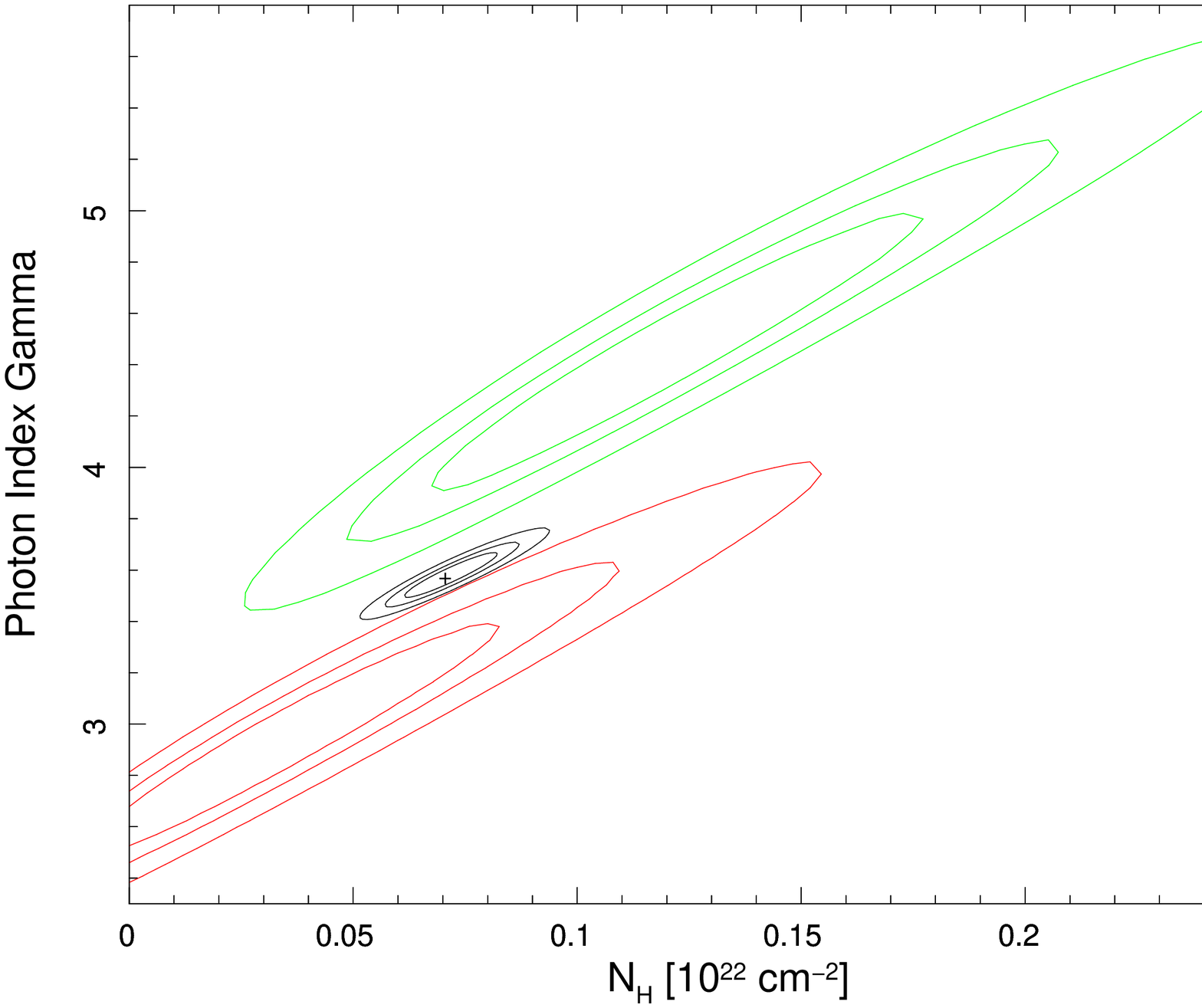}
\caption{\label{rxj0148_contour} Comparison between the spectra of the 2005 May, 2005
December 09, and December 20/21 observations. The left panel displays the 
spectra with the average of the 2005 May spectra (segment 002-006) = black,
December 09 (segment 009) = red, and December 20/21 (segment 010/011) = green.
The right panel shows the 
contour plot between the intrinsic column 
density $N_{\rm H, intr}$ and the soft X-ray 
photon index $\Gamma$ = \axs +1 of the XRT spectra. The colors of the contours
refer to the same segments as in the left panel. 
 }
\end{figure*}

A simple way to explain a hardening in the spectrum with
decreasing observed X-ray flux is with a cold absorber cloud in the
line of sight. Variable absorbers in AGN are often observed in Seyfert
galaxies, e.g. the Seyfert 2 sample of \citet{risaliti02}, NGC 1365
\citep{risaliti04}, NGC 4388 \citep{elvis04}, the Seyfert 1.8 galaxy
NGC 3786 \citep{kom97b} the Seyfert 1.5 galaxies NGC 4151
\citep{puccetti04} and NGC 3227 \citep{kom97a}, or 1H0419--577
\citep{pounds04}. A variable cold absorber could also, in part,
provide a plausible explanation for the spectral variability between
the 2005 May and 2005 December 09 observation. As listed in Table
\ref{xrt_fit}, we fitted an absorbed broken power law to the 2005
December 09 data, where all parameters were fixed to those determined
from the May 2005 observations, except for the absorption column
density and the normalization, which were allowed to vary. This
provided a best-fit column density \nh=8.1\plm1.4$\times 10^{20}$
cm$^{-2}$, although there were strong residuals below 0.5
keV. Although NLS1s often resemble AGN with only minor intrinsic
absorption, this is not a true picture in general
\citep[e.g. ][]{gru98b, grupe04c}. A different result is found if all
parameters are left to vary. As listed in Table \ref{xrt_fit} and
shown in Figure \ref{rxj0148_contour}, \nh~actually becomes consistent
with the Galactic value and the soft X-ray spectral index
\axs~flattens out. However, in a fitting routine like {\it XSPEC},
\nh~and the spectral index are not independent parameters. A larger
value of the absorption column density \nh~will result in a steeper
spectral index, and vice-versa. Given the column density observed in
December 20/21, which was again in the order of 10$^{21}$ cm$^{-2}$,
we can conclude that the spectral change seen between the 2005 May and
2005 December 09 observations is most likely due to an increase in the
absorber column density.  A soft X-ray spectrum fitted by a spectral
model can mimic a low column density as shown by e.g. \citet{puch95}
and \citet{gru98a}, even though the real column density is much
larger.

A softening with decreasing X-ray flux can be cause by several
processes such as a change in the accretion disk corona
\citep[e.g. ][]{gru00} or the presence of a variable ionized absorber
\citep[e.g. ][]{kom00}.  Another possibility is the presence of a
partial covering absorber as discussed for e.g. RX J2217.9--5941
\citep{gru04b}, 1H 0707--495 \citep{gallo04b, tanaka04}, and Mkn 1239
\citep{grupe04c}.  As an alternative, \citet{fabian04} discussed the
variability observed in 1H 0707--495 in the context of X-ray
reflection on an ionized disk. Of all these models only the partial
covering absorber model yields reasonable results. Interestingly, the
coverage fraction observed in 2005 May, 2005 December 09, and December
20/21 observation remains the same, at around $f$=0.8. The column
density of the partial covering absorber follows the same trend as the
cold absorber column density, suggesting that it is at a low value
during the December 09 observation.

The soft X-ray slope $\alpha_{\rm X, soft}$=2.58$^{+0.15}_{-0.12}$ is
rather steep even for a NLS1. The mean soft X-ray slope for the sample
of 51 NLS1s from \citet{gru04a} is \ax=1.96 with a standard deviation
$\sigma$=0.41, and \ax=2.1 for the sample taken from
\citet{bol96}. However, the hard X-ray spectral slope of
\ax=0.96$^{+0.16}_{-0.12}$ is slightly flatter than that found in the
sample of NLS1s from \citet{lei99b}, who found a mean hard X-ray slope
of \ax=1.19\plm0.10 and the sample from \citet{brandt97} with
\ax=1.15. This is in better agreement with the values found for BLS1,
for which \citet{lei99b} found \ax=0.78\plm0.11 and \citet{brandt97}
found \ax=0.87. A possible explanation for this `discrepancy' is that
whereas the soft X-ray spectral slope is driven by the Eddington
ratio $L/L_{\rm Edd}$, the hard X-ray spectral slope is more
dependent on the black hole mass. The Eddington ratio $L/L_{\rm Edd}$
is one of the highest in the sample from \citet{gru04a}, with
$L/L_{\rm Edd}$ = 4.  As shown by \citet{gru04c,mathur05a, mathur05b},
NLS1s with a high Eddington ratio $L/L_{\rm Edd}$ deviate
significantly from the $M_{\rm BH}$ - stellar velocity
dispersion $\sigma$ relation \citep[e.g. ][]{geb00a, ferr00, tre03}.
With a FWHM([OIII])=700\plm500 and a black hole mass in the order of a
few 10$^7$\msun~ \rxj\ shows one of the most extreme deviations from
the \citet{tre03} $M_{\rm BH}-\sigma$ relation.
 
\rxj~also shows a variation in its optical-to-X-ray spectral slope
\aox.  While during the 2005 December 09 observations \aox=1.5, during
the December 20/21 observations \rxj~became X-ray weak with \aox=1.81.
Similar changes in \aox~ have also been recently reported by
\citet{gallo06}.  Changes like these can explain in part the
large scatter seen in the \aox~diagrams of e.g. \citet{yuan98a,
  yuan98b} and \citet{strateva05}.

The \swift~ observations of RX J0148.3--2758 have shown the great
potential of \swift\ for AGN science. The X-ray light curves are
highly variable, in particular those of NLS1s, and require long-term
coverage in a range of wavelengths.  Due to its low-earth orbit,
\swift\ is very similar to {\it ROSAT} and {\it ASCA}, but has also the added
advantage of being a multi-wavelength observatory. Our study has shown
the importance of simultaneous UV and X-ray observations over a
time-span of days, and \swift\ is the only observatory that can obtain
such observations. Our observations of \rxj~utilize the
multi-wavelength capabilities of \swift\ as well as its flexible
observing scheduling. The simultaneous observations in the UVOT and
XRT allow us to measure the X-ray loudness \aox~directly without
assuming any optical/UV spectral slopes. We are also able to measure
the total power in the Big Blue Bump and therefore the bolometric
luminosity directly. The strong change in its spectrum between the
2005 May and 2005 December 09 observations prompted us to execute
further observations, which took place a few days later, on December
20/21. These additional observations allowed us to observe a hardening
and a softening in the same source. Based on this interesting spectral
behavior, we plan to continue observing \rxj~with \swift.

\acknowledgments

We would like to thank the whole \swift-team for making this observation
possible, especially the \swift\ science planners Jamie Kennea, Sally
Hunsberger, Claudio Pagani, Judy Racusin and Antonino Cucchiara
for scheduling RX J0148.3--2758 for such a long observing time, and Neil Gehrels
for approving the ToO observations of 2005-December 20/21.
We would also like to thank Marco Ajello and Jochen Greiner (MPE) 
and Jack Tueller (GSFC) for checking
the BAT pointed and survey data for any detection of RX J0148.3--2758.
We also thank the anonymous referee for valuable comments and suggestions to
improve this paper.
This research has made use of the NASA/IPAC Extra-galactic
Database (NED) which is operated by the Jet Propulsion Laboratory,
Caltech, under contract with the National Aeronautics and Space
Administration.
 This research was supported by NASA contract NAS5-00136 (D.G., D.B., \& J.N.).

\clearpage

%% Use the figure environment and \plotone or \plottwo to include

%% figures and captions in your electronic submission.

\clearpage

\begin{deluxetable}{llccrc}
%\tabletypesize{\scriptsize}
\tablecaption{Observation log of RX J0148.3--2758
\label{obs_log}}
\tablewidth{0pt}
\tablehead{
\colhead{Observatory} &
\colhead{observation} & \colhead{T-start\tablenotemark{1}} & 
\colhead{T-stop\tablenotemark{1}} &
\colhead{$\rm T_{exp}$\tablenotemark{2}} &
\colhead{log $L_X$\tablenotemark{3}}
} 
\startdata
\swift & segment 002\tablenotemark{4} & 2005-05-06 00:05 & 2005-05-06 24:00 & 7637  & 37.85\tablenotemark{5} \\
       & segment 003\tablenotemark{4} & 2005-05-07 00:05 & 2005-05-09 22:58 & 21966  \\
       & segment 004\tablenotemark{4} & 2005-05-11 00:27 & 2005-05-11 23:11 & 8840   \\
       & segment 006\tablenotemark{4} & 2005-05-13 00:45 & 2005-05-13 10:26 & 3132   \\
       & segment 008\tablenotemark{4} & 2005-12-07 00:34 & 2005-12-07 10:35 &  822 & 37.83   \\
       & segment 009\tablenotemark{4} & 2005-12-09 12:17 & 2005-12-09 23:54 & 6346 & 37.27 \\
       & segment 010\tablenotemark{4} & 2005-12-20 13:40 & 2005-12-20 23:32 & 5110 & 37.26\\
       & segment 011\tablenotemark{4} & 2005-12-21 01:01 & 2005-12-21 14:00 & 3506 & 37.21 \\
{\it ASCA}   & SIS 0, 1    & 1997-07-11 21:11 & 1997-07-13 03:07 & 33243 & 36.92 \\
       & GIS 2, 3    & 1997-07-11 21:11 & 1997-07-13 03:07 & 36381 \\             
{\it ROSAT}  & pointed PSPC & 1992-07-09 09:54 & 1992-07-10 01:16 & 6652 & 37.79 \\
{\it ROSAT}  & RASS        & 1990-07-15 15:26 & 1990-07-16 07:28 &    89 & 37.87 \\
       &             & 1990-12-28 01:11 & 1990-12-29 07:37 &   140 & 38.05 \\
       &             & 1991-01-15 09:24 & 1991-01-17 07:49 &   274 & 38.00 \\
\enddata

\tablenotetext{1}{Start and End times are given in UT}
\tablenotetext{2}{Observing time given in s}
\tablenotetext{3}{Rest-frame 0.2-2.0 keV luminosity given in units of W}
\tablenotetext{4}{The term 'segment' is from the way Swift is scheduled. Swift is scheduled on
a day by day basis. A segment contains the observations of a source of a single
day (except during the weekends when \swift\ is scheduled for 3 days)}
\tablenotetext{5}{Average luminosity of segments 002-006}
\end{deluxetable}
\clearpage

\begin{deluxetable}{lcccccccr}
\tabletypesize{\scriptsize}
\tablecaption{Spectral parameters of the fits to the \swift-XRT spectra of RX J0148.3--2758
\label{xrt_fit}}
\tablewidth{0pt}
\tablehead{
 &  & 
\colhead{$N_{\rm H, Gal}$} & 
\colhead{$N_{\rm H, intr}$} &  & 
\colhead{kT} & \colhead{$E_{\rm break}$} & &   \\
\colhead{\rb{Obs. Date}} & \colhead{\rb{Model\tablenotemark{1}}} & 
\colhead{10$^{20}$ cm$^{-2}$} & 
\colhead{10$^{20}$ cm$^{-2}$} & 
\colhead{\rb{$\alpha_{\rm X, soft}$\tablenotemark{2}}} & 
\colhead{eV} & \colhead{KeV} &
\colhead{\rb{$\alpha_{\rm X, hard}$}} & \colhead{\rb{$\chi^2/\nu$}}  
} 
\startdata
May 06  & 1 &  1.50 (fix) & --- & 1.84\plm0.07 & ---     & ---	    & ---	  & 113/68 \\
& 1 & 3.33$^{+0.14}_{-0.12}$ & --- & 2.03$^{+0.16}_{-0.15}$ & ---     & ---	& ---	      & 107/67 \\
& 1 & 1.50 (fix) &3.01$^{+0.22}_{-0.19}$ & 2.08$^{+0.19}_{-0.17}$ & --- & --- & --- & 106/67 \\
& 2 & 1.50 (fix)  & --- & ---		& 126$^{+8}_{-11}$ & ---	& 1.27$^{+0.30}_{-0.21}$ & 83/65 \\ 
& 3 & 4.80\plm1.16 & --- & 2.28\plm0.14 & ---	& 1.87\plm0.25 & 0.95\plm0.24 & 83/65  \\
& 4 & 1.50 (fix)  &  7.95$^{+0.25}_{-0.34}$ & 2.59$^{+0.21}_{-0.29}$ &
    --- & 1.80$^{+0.16}_{-0.19}$ & 0.85$^{+0.35}_{-0.23}$ & 71/64 \\

May 07 & 1 &  1.50 (fix) & --- & 1.82\plm0.03 & ---     & ---	    & ---	  & 268/130 \\
 & 1 & 2.77\plm0.05 & --- & 1.96\plm0.05 & ---	 & ---  	 & ---         & 258/129 \\
 & 1 & 1.50 (fix0) & 2.34$^{+1.17}_{-1.09}$ & 2.02$^{+0.11}_{-0.10}$ & --- & --- & --- & 254/129 \\
 & 2 & 1.50 (fix)  & --- & ---  	 & 118$^{+5}_{-7}$ & ---	 & 1.23$^{+0.14}_{-0.13}$ & 179/127 \\ 
 & 3 & 4.85$^{+0.93}_{-0.67}$  & --- &  2.30$^{+0.12}_{-0.11}$ & ---	 & 1.72$^{+0.15}_{-0.16}$ & 
 1.01$^{+0.18}_{-0.17}$ & 174/126  \\
   & 4 & 1.50 (fix) &  7.61$^{+0.20}_{-0.17}$ & 2.61$^{+0.19}_{-0.16}$ &
    --- & 1.61\plm0.14 & 1.00$^{+0.19}_{-0.13}$ & 146/126 \\

May 11  & 1 &  1.50 (fix) & --- & 1.93\plm0.05 & ---     & ---	    & ---	  & 155/67 \\
   & 1 & 3.88\plm0.09 & --- & 2.13\plm0.10 & ---     & ---	    & ---	  & 145/66 \\
   & 2 & 1.50 (fix)  & --- & ---	    & 120\plm7 & ---	    & 0.87$^{+0.14}_{-0.23}$ & 84/65 \\ 
   & 3 & 6.18$^{+1.81}_{-1.56}$ & --- & 2.53$^{+0.25}_{-0.20}$ & ---     & 1.75$^{+0.25}_{-0.30}$ & 
 0.69$^{+0.42}_{-0.36}$ & 98/63  \\
   & 4 & 1.50 (fix)  & 7.88$^{+0.36}_{-0.41}$ & 2.75$^{+0.32}_{-0.41}$ 
    & --- & 1.70$^{+0.33}_{-0.18}$ & 0.68$^{+0.33}_{-0.32}$ & 92/63 \\

May 13 & 1 &  1.50 (fix) & --- & 1.83\plm0.10 & ---     & ---	    & ---	  & 27/36 \\
   & 1 & 1.60$^{+0.17}_{-0.15}$ & --- & 1.84$x
 ^{+0.20}_{-0.18}$ & ---     & ---	    & ---	  & 27/35 \\
   & 2 & 1.50 (fix)  & --- &  ---	    & 108$^{+108}_{-102}$ & ---	     & 1.64$^{+0.25}_{-0.26}$ & 25/34 \\ 
   & 3 & 1.50 (fix)  & --- & 1.93\plm0.08 & ---     & 1.70 (fix) & 1.37\plm0.23 & 23/34  \\
   & 3 & 3.01$^{+1.80}_{-1.60}$ & --- & 2.10$^{+0.21}_{-0.19}$ & --- & 1.42\plm0.50 & 1.47\plm0.30 & 24/33 \\
   & 4 & 1.50 (fix)  & 3.42$^{+0.33}_{-0.11}$ & 2.24\plm0.16 & --- &
    1.40$^{+1.02}_{-3.24}$ & 1.47$^{+1.90}_{-3.11}$ & 23/32 \\
     
May 06-13\tablenotemark{3}  & 1 &  1.50 (fix) & --- & 1.84\plm0.03 & ---     & ---	    & ---	  & 567/304 \\
   & 1 & 2.92$^{+0.55}_{-0.53}$ & --- & 2.00\plm0.07 & ---     & ---	   & ---	 & 546/303 \\
   & 1 &  1.50 (fix) & 2.42$^{+0.86}_{-0.82}$ & 2.05\plm0.08 & ---     & ---	   & ---	 & 541/303 \\
   & 2 & 1.50 (fix)  & --- & ---	   & 119\plm4 & ---	   & 1.23$^{+0.04}_{-0.06}$ & 386/299 \\ 
   & 3 & 4.96$^{+0.68}_{-0.46}$ & --- & 2.33\plm0.08 & ---     & 1.76\plm0.11 & 0.96\plm0.13 & 386/300 \\
   & 4 & 1.50 (fix) & 7.15$^{+1.48}_{-1.21}$ & 2.58$^{+0.15}_{-0.12}$ &
    --- & 1.68$^{+0.12}_{-0.14}$ & 0.96$^{+0.16}_{-0.14}$ & 347/301 \\
    
December 07  & 1 & 1.50 (fix) & --- & 2.17$^{+0.27}_{-0.25}$ & --- & --- & --- & 12/7 \\
   & 2 & 1.50 (fix) & --- & --- & 104\plm16 & --- & 1.36$^{+1.21}_{-1.36}$ & 10/6  \\
   & 3 & 1.50 (fix) & --- & 2.26\plm0.32 & --- & 1.07\plm1.39 & 1.86\plm0.80 & 11/6 \\

December 09  & 1 & 1.50 (fix) & --- & 1.48$^{+0.10}_{-0.09}$ & --- & --- & --- & 30/37 \\
  & 2 & 1.50 (fix) & --- & --- & 100\plm20 & --- & 1.29$^{+0.22}_{-0.11}$ & 27/35 \\
  & 3 & 1.50 (fix) & --- & 1.60$^{+0.38}_{-0.15}$ & --- & 1.28$^{+4.30}_{-0.85}$ & 1.26$^{+0.29}_{-5.26}$ & 29/34 \\
  & 3 & 8.14$^{+1.49}_{-1.36}$ & --- & 2.33 (fix) & --- & 1.76 (fix) & 0.96 (fix) & 48/37 \\
  & 4 & 1.50 (fix) & 2.92$^{+6.07}_{-2.92}$ & 1.93$^{+0.58}_{-0.42}$ & --- & 1.16$^{+0.30}_{-0.19}$ & 1.27$^{+0.27}_{-0.15}$ & 25/34 \\
December 07+09\tablenotemark{3} & 1 & 1.50 (fix) & --- & 1.57\plm0.05 & --- & --- & --- & 60/47 \\
   & 2 & 1.50 (fix) & --- &  --- & 103$^{+23}_{-19}$ & --- & 1.23$^{+0.23}_{-0.12}$ & 38/43 \\
   & 3 & 2.27$^{+3.85}_{-1.80}$ & --- & 1.73$^{+0.64}_{-0.23}$ & --- & 2.01$^{+1.88}_{-0.65}$ & 
  1.29$^{+0.41}_{-0.71}$ & 54/43 \\
   & 4 & 1.50 (fix) & 6.93$^{+7.91}_{-5.96}$ & 2.45$^{+0.79}_{-0.73}$ & --- & 1.08$^{+1.10}_{-0.15}$ & 
  1.34$^{+0.23}_{-0.61}$ & 41/42 \\

December 20  & 1 & 1.50 (fix) & --- & 2.07$^{+0.21}_{-0.20}$ & --- & --- & --- & 22/12 \\
    & 4 & 1.50 (fix) & 2.86$^{+6.80}_{-2.80}$ & 2.47$^{+0.62}_{-0.42}$ & --- & 1.68 (fix) & 0.96 (fix) & 18/11 \\
 
 December 21 & 1 & 1.50 (fix) & --- &  1.96$^{+0.16}_{-0.15}$ & --- & --- & --- & 29/17 \\
 & 1 & 4.44$^{+3.25}_{-2.67}$ & --- & 2.33$^{+0.43}_{-0.37}$ & --- & --- & --- & 26/16 \\
 & 2 & 1.50 (fix) & ---  & --- & 110\plm14 & --- & 0.92$^{+0.37}_{-0.56}$ & 16/15 \\
 & 4 & 1.50 (fix) & 12.42$^{+9.91}_{-6.72}$ & 3.38$^{+1.00}_{-0.73}$ & --- & 1.30$^{+0.43}_{-0.26}$ & 0.98$^{+0.73}_{-0.95}$ & 12/14 \\
 
 December 20+21\tablenotemark{3} & 1  & 1.50 (fix) & --- & 2.00$^{+0.13}_{-0.12}$ & --- & --- & --- & 52/30 \\
 & 1 & 3.30$^{+2.42}_{-2.10}$ & --- & 2.22$^{+0.31}_{-0.21}$ & --- & --- & --- & 57/29 \\
 & 2 & 1.50 (fix) & --- & --- & 101\plm13 & --- & 1.13$^{+0.37}_{-0.39}$ & 33/27 \\
 & 3& 7.80$^{+4.31}_{-3.60}$ & --- & 3.05$^{+0.70}_{-0.55}$ & --- & 1.20$^{+0.49}_{-0.20}$ & 1.28$^{+0.44}_{-0.73}$ & 32/27 \\
 & 4 & 1.50 (fix) & 11.61$^{+7.49}_{-5.73}$ & 3.41$^{+0.78}_{-0.64}$ & --- &  1.17$^{+0.34}_{-0.17}$ & 1.32$^{+0.45}_{-0.52}$ & 30/27 
   
\enddata

\tablenotetext{1}{Model fit to the data: 1) Single power law with Galactic absorption; 2) Blackbody plus power law with
Galactic absorption; 3) Broken power law with Galactic absorption; 4) Broken power law with Galactic absorption and intrinsic
absorption at z=0.121}
\tablenotetext{2}{This spectral slope also refers to the 0.3-10.0 keV slope in
case only a single power law has been used.}
\tablenotetext{3}{Simultaneous fits in {\it XSPEC} }
\end{deluxetable}

\clearpage

\begin{deluxetable}{lccccccccr}
\tabletypesize{\scriptsize}
\tablecaption{Spectral parameters of the fits to the {\it ROSAT} and {\it ASCA}  spectra of RX J0148.3--2758
\label{rosat_fit}}
\tablewidth{0pt}
\tablehead{
\colhead{Mission} &
\colhead{observation} & \colhead{Model\tablenotemark{1}} & 
\colhead{$N_{\rm H, Gal}$\tablenotemark{2}} & 
\colhead{$N_{\rm H, intr}$\tablenotemark{2}} & 
\colhead{$\alpha_{\rm X, soft}$\tablenotemark{3}} & 
\colhead{kT\tablenotemark{4}} & \colhead{$E_{\rm break}$\tablenotemark{5}} &
\colhead{$\alpha_{\rm X, hard}$} & \colhead{$\chi^2/\nu$}  
} 
\startdata
{\it ASCA} & 
    & 1 &  1.50 (fix) & --- & 1.48$^{+0.09}_{-0.08}$ & ---     & ---         & ---         & 254/237 \\
  & & 2 & 1.50 (fix)  & --- & ---         & 127\plm16 & ---         & 1.10$^{+0.13}_{-0.14}$ & 208/235 \\ 
  & & 3 & 1.50 (fix) & --- & 2.03$^{+0.23}_{-0.20}$ & ---     & 1.36$^{+0.16}_{-0.19}$ & 
    1.11$^{+0.16}_{-0.19}$ & 210/235 \\
{\it ROSAT} & RASS & 1 &  1.50 (fix) & --- & 2.12\plm0.11 & ---     & ---         & ---         & 30/23 \\
  & & 1 & 2.54\plm0.82 & --- & 2.62\plm0.30 & ---     & ---         & ---         & 21/22 \\

{\it ROSAT} & po & 1 &  1.50 (fix) & --- & 1.88\plm0.03 & ---     & ---         & ---         & 178/110 \\
  & & 1 & 2.35\plm0.22 & --- & 2.25\plm0.08 & ---     & ---         & ---         & 111/110

\enddata

\tablenotetext{1}{Model fit to the data: 1) Single power law with Galactic absorption; 2) Blackbody plus power law with
Galactic absorption; 3) Broken power law with Galactic absorption; 4) Broken power law with Galactic absorption and intrinsic
absorption at z=0.121}
\tablenotetext{2}{Column density $N_{\rm H}$ given in units of 10$^{20}$
cm$^{-2}$}
\tablenotetext{3}{This spectral slope also refers to the 0.3-10.0 keV slope in
case only a single power law has been used.}
\tablenotetext{4}{kT in units of eV}
\tablenotetext{5}{Broken Power law break energy $E_{\rm break}$ in units of keV}

\end{deluxetable}

\begin{deluxetable}{lcccccccc}
\tabletypesize{\scriptsize}
\tablecaption{UVOT photometry from the co-added images of RX J0148.3--2758
\label{uvot_photometry}}
\tablewidth{0pt}
\tablehead{& \multicolumn{2}{c}{Segment 008} & \multicolumn{2}{c}{Segment 009} 
& \multicolumn{2}{c}{Segment 010} & \multicolumn{2}{c}{Segment 011} \\ 
\colhead{\rb{Filter}} &
\colhead{Mag} & \colhead{Flux\tablenotemark{1}} & 
\colhead{Mag} & \colhead{Flux\tablenotemark{1}} & 
\colhead{Mag} & \colhead{Flux\tablenotemark{1}} & 
\colhead{Mag} & \colhead{Flux\tablenotemark{1}}  
}
\startdata
V    & 15.358\plm0.062 & 26.90\plm1.18 & 15.350\plm0.018 & 27.10\plm0.35 & 15.321\plm0.020 & 27.84\plm0.39 & 15.299\plm0.017 & 28.41\plm0.34 \\
B    & ---             & ---           & 15.645\plm0.015 & 33.45\plm0.41 & 15.647\plm0.017 & 33.37\plm0.39 & 15.652\plm0.088 & 33.24\plm2.40 \\           
U    & 14.625\plm0.042 & 47.19\plm1.62 & 14.482\plm0.011 & 53.80\plm0.49 & 14.480\plm0.013 & 53.91\plm0.55 & 14.523\plm0.030 & 51.81\plm1.27 \\
UVW1 & 14.531\plm0.031 & 64.62\plm1.52 & 14.485\plm0.009 & 67.39\plm0.49 & 14.471\plm0.011 & 68.29\plm0.56 & 14.425\plm0.024 & 71.25\plm1.32 \\
UVM2 & 14.701\plm0.031 & 74.77\plm1.71 & 14.602\plm0.010 & 81.85\plm0.57 & 14.603\plm0.011 & 81.78\plm0.64 & 14.647\plm0.024 & 78.57\plm1.37 \\
UVW2 & 14.598\plm0.018 & 112.15\plm1.46 & 14.537\plm0.007 & 118.60\plm0.59 & 14.530\plm0.008 & 119.39\plm0.66 & 14.533\plm0.006 & 119.11\plm0.56
\enddata

\tablenotetext{1}{The fluxes are given in units of 10$^{-19}$ W m$^{-2}$ \AA$^{-1}$.}

\end{deluxetable}

\begin{deluxetable}{lcccccccc}
\tabletypesize{\scriptsize}
\tablecaption{UVOT photometry from the co-added images of the four comparison stars
The light curves of these stars is shown in Figure\,\ref{rxj0148_uvot_stars}.
\label{stars}}
\tablewidth{0pt}
\tablehead{\colhead{Star} & \colhead{segment} 
& \colhead{V} 
& \colhead{B} 
& \colhead{U} 
& \colhead{UVW1} 
& \colhead{UVM2} 
& \colhead{UVW2} 
}
\startdata
1 & 008 & 13.102\plm0.036 & ---             & 13.695\plm0.038 & 15.063\plm0.041 & 16.842\plm0.087 & 16.804\plm0.050 \\
  & 009 & 13.072\plm0.011 & 13.638\plm0.018 & 13.600\plm0.010 & 15.058\plm0.012 & 16.667\plm0.024 & 16.729\plm0.017 \\
  & 010 & 13.074\plm0.012 & 13.634\plm0.020 & 13.599\plm0.012 & 15.046\plm0.013 & 16.670\plm0.027 & 16.728\plm0.020 \\
  & 011 & 13.074\plm0.010 & 13.676\plm0.010 & 13.600\plm0.028 & 15.026\plm0.031 & 16.672\plm0.061 & 16.707\plm0.016 \\
2 & 008	& 13.79\plm0.037  & ---	            & 14.29\plm0.040  & 15.56\plm0.056	& 16.96\plm0.100  & 17.20\plm0.063 \\
  & 009	& 13.81\plm0.011  & 14.38\plm0.013  & 14.20\plm0.011  & 15.60\plm0.016  & 16.96\plm0.029  & 17.20\plm0.023 \\
  & 010	& 14.02\plm0.013  & 14.62\plm0.014  & 14.42\plm0.012  & 15.78\plm0.019  & 17.17\plm0.036  & 17.40\plm0.028 \\
  & 011	& 14.00\plm0.011  & ---             & ---             & ---             & ---             & 17.36\plm0.024 \\
3 & 008	& 15.09\plm0.057  & ---	            & 15.57\plm0.061  & 16.70\plm0.114  & 18.90\plm0.382  & 18.41\plm0.157 \\
  & 009	& 15.09\plm0.016  & 15.66\plm0.015  & 15.50\plm0.015  & 16.87\plm0.032  & 18.14\plm0.060  & 18.50\plm0.053 \\
  & 010	& 15.05\plm0.018  & 15.63\plm0.017  & 15.55\plm0.017  & 16.84\plm0.035  & 18.20\plm0.069  & 18.50\plm0.061 \\
  & 011	& 15.08\plm0.016  & 15.74\plm0.092  & 15.54\plm0.042  & 16.80\plm0.080  & 18.27\plm0.160  & 18.41\plm0.048 \\
4 & 008	& 16.77\plm0.162  & ---	            & 16.89\plm0.126  & 17.45\plm0.196  & 18.34\plm0.249  & 18.13\plm0.120 \\
  & 009 & 16.70\plm0.038  & 17.02\plm0.028  & 16.84\plm0.023  & 17.52\plm0.047  & 18.09\plm0.056  & 18.38\plm0.048 \\
  & 010	& 16.69\plm0.043  & 17.02\plm0.031  & 16.85\plm0.033  & 17.53\plm0.053	& 18.12\plm0.062  & 18.40\plm0.053 \\
  & 011	& 16.68\plm0.036  & 17.16\plm0.174  & 16.74\plm0.073  & 17.45\plm0.116  & 18.02\plm0.131  & 18.33\plm0.044 
\enddata

\end{deluxetable}

\begin{deluxetable}{lrrrccl}
%\tabletypesize{\scriptsize}
\tablecaption{\label{sed_data}Measurements of the spectral energy distribution
 shown in Figure \ref{rxj0148_sed}. 
}
\tablewidth{0pt} 
\tablehead{
\colhead{Observatory} &
\colhead{Filter} & \colhead{$\lambda_{\rm c}$\tablenotemark{1}} & 
\colhead{log $\nu$ [Hz]} &
\colhead{Magnitude\tablenotemark{2}} &
\colhead{$\nu L_{\nu}$\tablenotemark{3}} &
\colhead{Comments}
} 
\startdata
NVSS & 1.40 GHz & ---        & 9.146 & $<$1mJy & \tablenotemark{4} & \\
{\it IRAS} & 100$\mu$m & 100$\mu$m & 12.477 & 746\plm190 mJy & 7.41\plm1.89 & \\
     &  60$\mu$m &  60$\mu$m & 12.699 & 237\plm50  mJy & 3.98\plm0.83 & \\
     &  25$\mu$m &  25$\mu$m & 13.079 & 107\plm25  mJy & 4.27\plm1.00 & \\
     &  12$\mu$m &  12$\mu$m & 13.398 & 113\plm30  mJy & 9.55\plm2.54 & \\
2MASS & Ks       & 2.159$\mu$m & 14.143 & 12.250\plm0.026 & 3.85\plm0.05 & \\
      & H        & 1.662$\mu$m & 14.302 & 13.399\plm0.032 & 2.99\plm0.10 & \\
      & J        & 1.235$\mu$m & 14.385 & 14.214\plm0.025 & 2.67\plm0.07 & \\
UVOT  & V        & 5460\AA     & 14.740 & 15.37\plm0.02   & 5.08\plm0.25 & Segment 010\\
      & B        & 4340\AA     & 14.840 & 15.66\plm0.02   & 4.84\plm0.24 & Segment 010\\
      & U        & 3450\AA     & 14.939 & 14.49\plm0.01   & 6.22\plm0.31 & Segment 010\\
      & UVW1     & 2600\AA     & 15.062 & 14.49\plm0.01   & 5.88\plm0.30 & Segment 010 \\
      & UVM2     & 2200\AA     & 15.135 & 14.60\plm0.01   & 6.01\plm0.30 & Segment 010\\
      & UVW2     & 1930\AA     & 15.191 & 14.54\plm0.01   & 7.70\plm0.38 & Segment 010            
\enddata

\tablenotetext{1}{Central wavelength of the filter}
\tablenotetext{2}{For the NVSS and the {\it IRAS} data we give the flux densities in units of mJy. All others are given in units of
mag.}
\tablenotetext{3}{Observed Luminosities are  given in units of 10$^{37}$ W.}
\tablenotetext{4}{The upper limit of the NVSS observation is $\nu L_{1.4GHz}<4.7\times 10^{31}$ W.}
\end{deluxetable}

\end{document}